\documentclass{aa}
\usepackage{graphicx}

\begin{document}

\title{Systematic Effects in Measurement of Black Hole 
Masses by Emission-Line Reverberation of Active Galactic
Nuclei: Eddington Ratio and Inclination}

\author{Suzy Collin\inst{1}, Toshihiro Kawaguchi\inst{2}, 
Bradley M.\ Peterson\inst{3}, and Marianne Vestergaard\inst{4}}

\offprints{Suzy Collin (suzy.collin@obspm.fr)}

\institute{$^1$ LUTH, Observatoire de Paris, Section de
Meudon, F-92195 Meudon Cedex, France\\
$^2$  Optical and Infrared Division, NAOJ, Mitaka,
Tokyo 181-8588, Japan\\
$^3$ Department of Astronomy, The Ohio State University, 
140 West 18th Avenue, Columbus, OH 43210, USA \\
$^4$ Steward Observatory, University of Arizona, 933 N. Cherry Avenue, 
Tucson, AZ 85721, USA }

\titlerunning{Masses of Black Holes in AGNs}
\authorrunning{S. Collin, T. Kawaguchi, B.M. Peterson, \& M. Vestergaard}

\abstract{Scatter around the
relationship between central black hole masses in active galactic nuclei (AGNs) obtained by  reverberation-mapping methods and 
host-galaxy bulge velocity dispersion indicates that the
masses are uncertain typically by a factor of about three.}
{In this paper, we try to identify
the sources and systematics of this uncertainty.}
{We characterize the broad H$\beta$ emission-line
profiles by the ratio of their full-width at half maximum (FWHM)
to their line dispersion, i.e., the second moment of the
line profile. We use this parameter to separate the
reverberation-mapped AGNs into two populations,
the first with narrower H$\beta$ lines that tend to
have relatively extended wings, and the second
with broader lines that are relatively flat-topped.
The first population is characterized by
higher Eddington ratios than the second. Within
each population, we calibrate the black-hole mass scale
by comparison of the reverberation-based mass with
that predicted by the bulge velocity dispersion.
We also use the distribution of ratios of the
reverberation-based mass to the velocity-dispersion mass
prediction in a comparison with a ``generalized
thick disk'' model in order to see if inclination
can plausibly account for the observed distribution.}
{We find that the line dispersion is a less biased
parameter in general than FWHM for black hole mass estimation, although
we show that it is possible to empirically correct for
the bias introduced by using FWHM to characterize the
emission-line width.
We also argue that inclination effects are apparent only
in some small subset of the reverberation-based
mass measurements; it is primarily the objects with the
narrowest emission lines that seem to be most strongly affected.}
{Our principal conclusion is that the H$\beta$ profile is
sensitive primarily to Eddington ratio, but that inclination
effects play a role in some cases.}


\maketitle

\section{Introduction}

During the last twenty years, reverberation mapping 
(Blandford \& McKee 1982; Peterson 1993)
of the broad emission lines in active galactic nuclei (AGNs) has
been used to determine the size of the broad-line region
(BLR) in these objects. By combining the BLR size with
the emission-line Doppler width, it is possible to estimate
the mass of the central source, presumed to be a black hole (BH), as
\begin{equation}
M_{\rm BH}=f{\ R_{\rm BLR} \Delta V^2 \over G},
\label{eq-virial}
\end{equation}
where $R_{\rm BLR}$ is the size of the BLR, $\Delta V$
is the emission line width, and $f$ is 
a scale factor of order unity that
depends on the structure, kinematics, and inclination of the BLR. 
In the cases where multiple emission lines have been observed,
it is found that the higher-ionization lines tend to
have shorter response times, or lags $\tau$, and thus arise in
regions relatively closer to the central source than
the lower-ionization lines. Moreover, the lines with
shorter lags tend to be broader than those with longer
lags, generally consistent with the virial prediction
$\tau = R/c \propto \Delta V^{-2}$, providing a strong
argument that eq.\ (1) is valid
(Peterson \& Wandel 1999, 2000, Onken \& Peterson 2002;
Kollatschny 2003a). In addition,
AGNs show a relationship between BH mass and 
the host-galaxy bulge velocity dispersion
$\sigma_*$ (Gebhardt et al.\ 2000b; Ferrarese et al.\ 2001;
Onken et al.\ 2004; Nelson et al.\ 2004) that is
consistent with this same correlation, the
$M_{\rm BH}$--$\sigma_*$ relationship, that is observed
in quiescent galaxies (Ferrarese \& Merritt 2000;
Gebhardt et al.\ 2000a; Tremaine et al.\ 2002).

Reverberation mapping has also confirmed the expected
relationship between the size of the BLR and the
luminosity of the AGN, which takes the form
\begin{equation}
R_{\rm BLR} \propto L^{\alpha},
\end{equation}
where $\alpha \approx 0.5$, but depends somewhat on
which luminosity measure is being used and possibly
also on which emission line is used to
determine $R_{\rm BLR}$ (Wandel, Peterson, \& Malkan 1999;
Kaspi et al.\ 2000, 2005; Bentz et al.\ 2006).
This radius--luminosity ($R$--$L$) relationship
is important not only because it tells us something
about the physics of the emission-line region, but
because it also affords an almost trivially simple
secondary mass indicator, since through eq.\ (2),
measurement of the
luminosity provides a surrogate for the otherwise
hard-to-measure BLR radius. It thus becomes possible to
easily estimate BH masses for large
samples of AGNs (Wandel et al.\ 1999;
Vestergaard 2002, 2004; McLure \& Jarvis 2002;
Kollmeier et al.\ 2006; Vestergaard \& Peterson 2006).
Such studies have shown, for example, that 
the BH masses of high-redshift quasars are very large (Vestergaard 2004; 
Netzer 2003). At the other extreme, very small BH masses have
been inferred for low-luminosity Seyferts (Greene \& Ho 2004; Barth et
al.\ 2005). It has also been found that application
of luminosity-scaling relationships indicate that
AGNs with small line widths, i.e., narrow-line Seyfert 1 (NLS1)
galaxies have relatively small BH masses for their luminosity.
It appears that the ratio of their bolometric luminosity
to the Eddington luminosity, which we hereafter refer to
as the ``Eddington ratio,'' is high, close in fact to unity.
Through accretion-disk modeling,
Collin et al.\ (2002), Kawaguchi (2003) and Collin \& Kawaguchi
(2004) have shown that NLS1s have accretion rates larger than the
Eddington value (although also see Williams, Mathur, \& Pogge 2004).

\begin{table*}
\begin{center}
\includegraphics{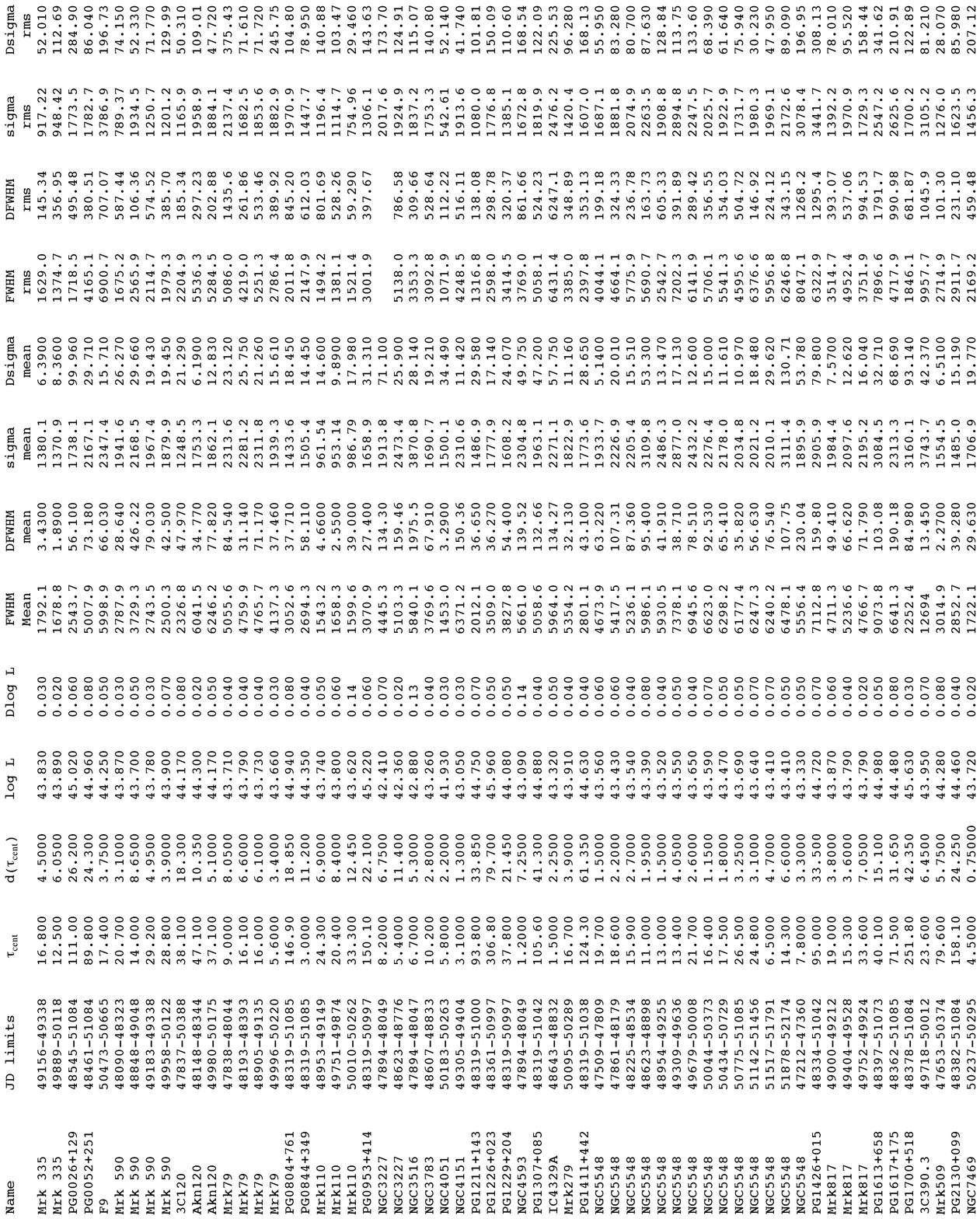}
\caption{Principal quantities of the reverberation mapped AGNs.}
\label{table1}
\end{center}
\end{table*}

These results have important physical and
cosmological consequences (cf.\ Yu \& Tremaine 2003; Kawaguchi et
al.\ 2004b). It is therefore essential to understand the
accuracy of the reverberation measurements (primary mass indicators) and
their calibration uncertainties as well as the scaling
relationships (secondary mass indicators) based on them,
especially when these are extrapolated beyond the
range over which they were determined, i.e., to much higher and 
much lower luminosities, to high redshifts, and to high and low Eddington
ratios. In principle, this could be achieved through comparison
with an independent primary or secondary mass indicator, 
though at the present time there is no other obvious
choice beyond the $M_{\rm BH}$--$\sigma_*$ relationship.
Indeed, the $M_{\rm BH}$--$\sigma_*$ relationship has been
employed in investigations of AGN BH masses in
two distinct ways, (a) by using measurements of $\sigma_*$
to infer a BH mass to compare with the
reverberation-based mass in an attempt to extract
information on the inclination of the BLR
(Wu \& Han 2001; Zhang \& Wu 2002), and 
(b) by normalizing the AGN $M_{\rm BH}$--$\sigma_*$ relationship
to that of quiescent galaxies in order to
calibrate the reverberation-based mass scale
by determining a statistical average value
for the scaling factor $\langle f \rangle$
(Onken et al.\ 2004).

Of course, it would be even more desirable to use 
reverberation-mapping techniques to obtain a velocity--delay map
that would reveal the kinematics and structure
of the BLR and lead to determination of the central
mass. However, various limitations of the reverberation data obtained
to date have precluded this (Horne et al.\ 2004),
but have nevertheless yielded mean
response times for emission lines and have enabled BH
mass estimates through eq.\ (1). This apparent simplication
entails a price in the accuracy to which the BH
mass can be measured and a number of ambiguities arise.
Among the more challenging are:
\begin{enumerate}
\item What line-width measure gives the most accurate
BH mass? The two commonly used measures are
the full width at half maximum (FWHM) and the
second moment of the emission-line profile
$\sigma_{\rm line}$, which we refer to as the 
``line dispersion\footnote{The line dispersion for an
emission line $\sigma_{\rm line}$ should not be confused
with the host galaxy bulge velocity disperion
$\sigma_*$.}.'' The ratio of these two measures varies
widely from one AGN to another and, as discussed below,
varies with other AGN spectral properties. Thus, depending
on which measure we select, we might be introducing a
bias in mass estimate that varies with other AGN parameters.
\item How does one evaluate and interpret the scaling
factor $f$? It is clear that $f$ should depend strongly on the inclination
for a flattened system (e.g., a rotating disk), but it
may vary with other AGN properties. Given that we are
characterizing the size and velocity dispersion of the BLR by single
numbers, we are subsuming a lot of our ignorance of AGN
structure into this single parameter, so interpretation
can be difficult and ambiguous.
\end{enumerate}

In this contribution, we will examine the systematics of
the line-width measures and attempt to identify possible
biases associated with the character of the emission-line
profiles. We will also, in the spirit of an exercise, 
assume that the H$\beta$-emitting component of the BLR 
is an azimuthally symmetric, relatively flat structure that 
is dominated by rotational motion (i.e., a Keplerian disk). We will
combine this with an isotropic component, a wind perhaps,  and investigate
the effects of disk inclination on the scaling factor $f$.
Our goal is to attempt to disentangle the different factors which
play a role in the determinimg line profile of AGNs, and therefore in the scale
factor and in the mass determination.

As we noted earlier, we subsume most of our ignorance about the
BLR structure and kinematics into the scale factor $f$
and separate it from the ``observable'' quantity in
eq.\ (1), which we refer to as the ``virial product,''
${\rm VP} = c\tau \Delta V^2/G$, which has units of mass
and differs from the actual BH mass only by the 
dimensionless factor $f$.
In this contribution, we will not concern ourselves with
uncertainties in the virial product, which are generally
$\sim 30$\%. It is the much larger uncertainty in $f$
that will concern us here.

In the next section, we study the relationship between FWHM and
$\sigma_{\rm line}$ and we show that the ratio of these quantities
varies strongly among the objects. We try to understand
the cause and the implications of these variations. By comparing the
virial products with the BH masses determined through the 
$M_{\rm BH}$--$\sigma_*$ 
relationship, we consider the relationships among the scale factor, the
line-width ratio ${\rm FWHM}/\sigma_{\rm line}$, and the Eddington ratio.  
In Section 3, we discuss the influence of the inclination 
of the BLR on the scale factor. Within the context of our
simple two-component model, we will argue
that some objects, specifically some of
those for which ${\rm FWHM}/\sigma_{\rm line}$ is small,
are likely seen nearly face-on geometry, and consequently their
BH masses could be significantly underestimated. 
In Section 4, we speculate on the physical conditions in the
BLR that could explain our results.
We conclude with a summary in Section 5.

\section{Normalization of Reverberation Masses: The Scale
Factor $f$}

The fundamental challenge that we need to address is how to
obtain from reverberation data masses that are accurate and
unbiased with respect to the various factors, such as inclination
of the BLR, that can affect the value of the virial product.
Given our poor understanding of the structure and kinematics of the
BLR we do not know a priori what these various factors might be.
We do know, however, that the broad H$\beta$ profiles 
show marked differences among AGNs, and it is certainly
true that the mass we adopt for a particular AGN will depend on
precisely how we characterize the width of the line. Thus, 
our approach will be necessarily empirical: we will examine various
ways of determining the line-width measure that is used as 
$\Delta V$ in eq.\ (1) and attempt to identify systematic effects 
or biases.

\subsection{Mean and RMS Spectra}

In computing reverberation-based masses, it is common
practice to take all of the individual spectra that were
measured to obtain the continuum and emission-line light
curves and construct ``mean'' and 
``root-mean-square (rms)'' spectra. The
advantage of the rms spectrum over the mean spectrum is
that it isolates the spectral components that are
actually varying, and automatically
removes constant or slowly varying components, such as
the narrow emission lines that arise on much larger
physical scales. The disadvantage of using the rms spectrum
is that it is often quite noisy as the amplitude of
spectral variability is usually rather low.
It is consequently not obvious whether it is better to measure 
the line width in the mean or the rms spectrum.
We will argue here that use of the mean spectrum for
line-width measurement gives results consistent with
line-width measurements based on rms spectra, provided
that one can account for contaminating features,
the narrow-line components being most important in
the case of the Balmer lines. This is important because
in using single-epoch spectra and scaling relationships
to estimate BH masses, the variable part of the
emission line cannot be isolated.

\subsection{Line-Width Measures: {\rm FWHM} and $\sigma_{\rm line}$}

We first consider the differences between the FWHM and the line
dispersion $\sigma_{\rm line}$. 
The FWHM is a zeroth-order moment of the profile 
and the line dispersion is a second-order moment that is relatively
more sensitive to the line wings and less sensitive to the line
core. It is traditional in AGN studies to use FWHM to characterize
line widths. For determination of BH masses,
Wandel et al.\ (1999) and Kaspi et al.\ (2000) used 
FWHM, but based on a suggestion by Fromerth \& Melia (2000),
Peterson et al. (2004) investigated use of the line
dispersion and found that it can be measured to higher
{\em precision} (i.e., with the smaller formal uncertainty)
than FWHM, especially in noisy spectra, and
that the virial relationship $\tau \propto \Delta V^{-2}$
is better reproduced with $\sigma_{\rm line}$ than with
FWHM. From the point of view of {\em accuracy}
(i.e., a measurement closest to the true value),
however, it is  
not clear which is the best line-width measure to use in
eq.\ (1), and it is that question we take up here.

The relationship between $\sigma_{\rm line}$ and FWHM depends on the
line profile: it is well-known, for example, that for a
Gaussian profile 
${\rm FWHM}/\sigma_{\rm line} = 2 (2 \ln 2)^{1/2} = 2.35$.
For a rectangular function,
${\rm FWHM}/\sigma_{\rm line} = 2\sqrt{3}=3.46$, and a triangular
function has 
${\rm FWHM}/\sigma_{\rm line} = \sqrt{6}=2.45.$ Similarly, 
for an edge-on rotating ring, 
${\rm FWHM}/\sigma_{\rm line} = 2\sqrt{2}=2.83.$ At the lower
extreme, ${\rm FWHM}/\sigma_{\rm line} \rightarrow 0$ for
a Lorentzian profile.

Peterson et al.\ (2004) provide measurements of FWHM and
$\sigma_{\rm line}$ from the rms spectra for all
the lines for which time-delay measurements are available
for virtually all of the reverberation-mapped AGNs
(see their Table 6). For the purpose of comparison, we
have measured the H$\beta$ line widths in the 
corresponding mean spectra, using the same algorithms and assumptions
of Peterson et al.\ (2004) and removing the 
narrow-line components whenever necessary. 
Also, the H$\beta$ profiles in the mean spectra often have
strong contamination in the long-wavelength wing, underneath the
[\ion{O}{III}]\,$\lambda\lambda4959,$ 5007 lines, by
\ion{Fe}{II} m42 emission. This contamination is much
weaker in the rms spectra because \ion{Fe}{II} emission
seems to be less variable than the Balmer lines
(cf.\ Vestergaard \& Peterson 2005). For this reason,
the $\sigma_{\rm line}$ measurements we use are based
on the short-wavelength side of the line, assuming 
that line is approximately symmetric.
These measurements, plus
the H$\beta$ rms spectrum measurements from Peterson et al.\ (2004),
are given in Table \ref{table1}, and these values will be
used throughout the rest of this paper. Columns (1) and (2)
identify the AGN and time interval (Julian Date $-$ 2400000)
during which the data were obtained, respectively.
Column (3) gives the cross-correlation function
centroid $\tau_{\rm cent}$, in days, which is our preferred
measure of the emission-line lag. Column (4) gives an estimate
of the uncertainty $\Delta \tau_{\rm cent}$, that is the root mean square 
of the usually slightly asymmetric upward and downward uncertainties,
again in days. Columns (5) give the logarithm of the 
mean optical luminosity in erg\,s$^{-1}$ based on the continuum measurements
made during the interval given in column (2),
and column (6) gives the rms variability of the luminosity
during the same interval.
The following columns give the values and uncertainties
all in km\,s$^{-1}$ in the rest frame of the AGN and corrected
for the spectrograph resolution,  for
FWHM in the mean spectrum (columns 7 and 8),
$\sigma_{\rm line}$ in the mean spectrum (columns 9 and 10),
FWHM in the rms spectrum (columns 11 and 12), and
$\sigma_{\rm line}$ in the rms spectrum (columns 13 and 14),
We note that the
optical luminosity has not been corrected for the host-galaxy
starlight contribution.
Therefore, the luminosity and Eddington ratios, especially for 
the less-luminous objects, are subject to overestimation 
on account of starlight contamination.
Some of the values in this Table will be
superceded by work in progress, but since the database is
constantly evolving, we have decided to proceed with the present data, as our
conclusions are not likely to change drastically by a few more
precise measurements.  

\begin{figure*}
\begin{center}
\includegraphics[width=\linewidth]{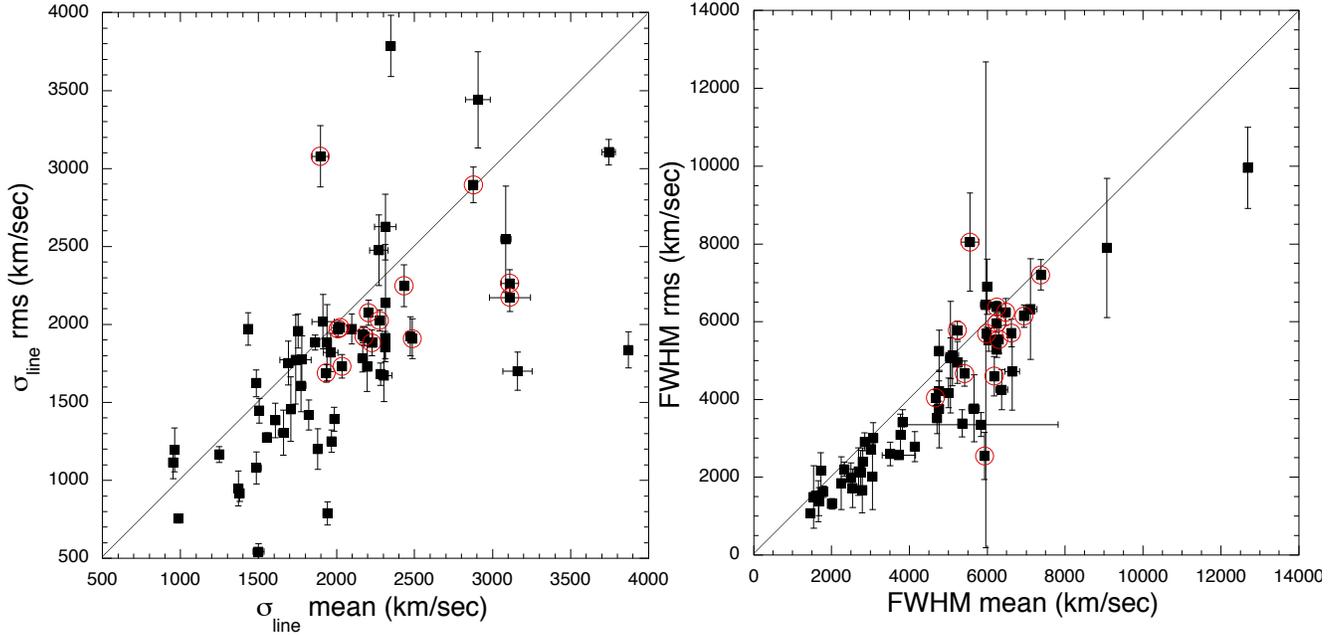}
\caption{The left panel shows $\sigma_{\rm line}$ in the rms spectrum
versus $\sigma_{\rm line}$ in the mean spectrum for H$\beta$ 
in all 35 datasets in Table 1. The right panel
shows the FWHM in the rms spectrum versus the FWHM in the mean
spectrum. The open circles mark the different datasets of 
NGC~5548. The straight line traces the diagonal. }
\label{fig-sig-FW-rms-vs-sig-FW-mean}
\end{center}
\end{figure*}

In Fig.\ \ref{fig-sig-FW-rms-vs-sig-FW-mean}, we compare
measurements of $\sigma_{\rm line}$ for H$\beta$
in the mean and rms spectra (left panel) and 
FWHM for H$\beta$ in the mean and rms spectra (right panel).
Both FWHM and $\sigma_{\rm line}$ measurements in the
rms spectra have much larger error bars
since the rms spectra tend to be noisy.
But we do see that for both line-width measures
the widths of the H$\beta$ line in the mean and
rms spectra are well correlated, though the lines
are typically $\sim20$\% broader in the mean spectra.
This is a well-documented phenomenon (e.g.,
Sergeev et al.\ 1999; Shapovalova et al.\ 2004).
It has been suggested by Shields, Ferland, \& Peterson (1995)
that the highest velocity gas in the BLR is optically
thin, and this could account for the lower level of
variability in the line wings.

\begin{figure*}
\begin{center}
\includegraphics[width=\linewidth]{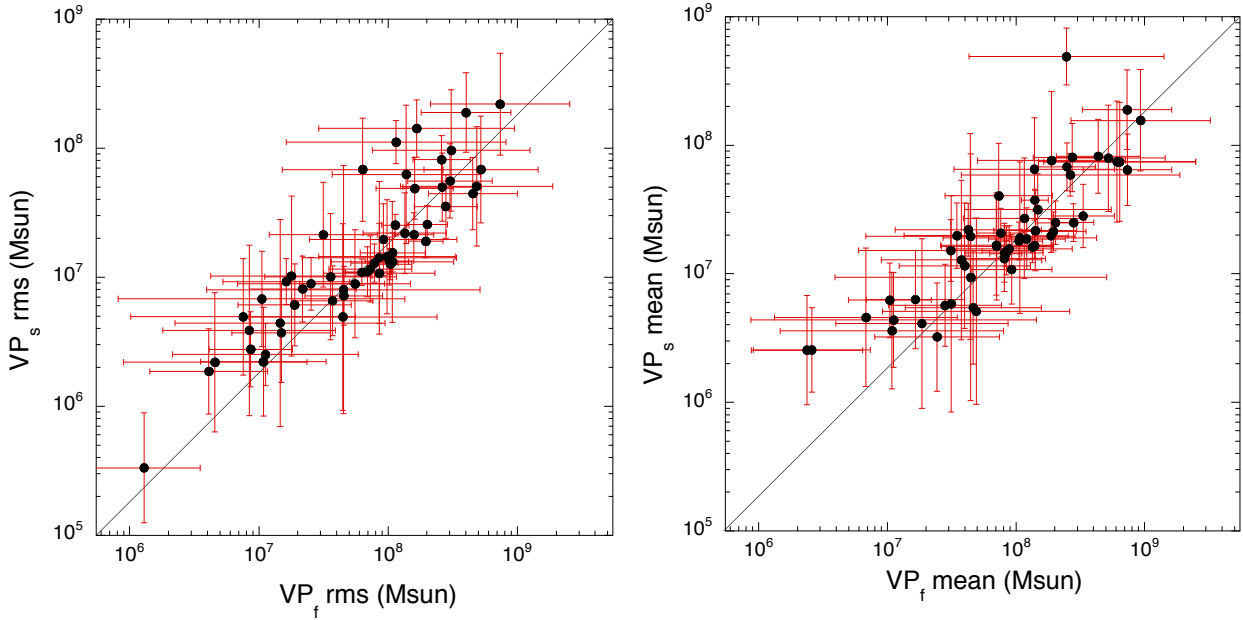}
\caption{Virial product based on $\sigma_{\rm line}$ (VP$_{\rm s}$),
versus the virial product based on the FWHM (VP$_{\rm f}$), for all
but four of the datasets (identified in the text)
in Table 1, based on rms spectra
(left panel) and  mean spectra (right panel).  The
straight lines correspond to a Gaussian profile, for which
${\rm VP}_{\rm s}= {\rm VP}_{\rm f}/5.52$.} 
\label{fig-VPsig-vs-VPFW}
\end{center}
\end{figure*}

In  Fig. \ref{fig-VPsig-vs-VPFW}, we compare the
virial product based on $\sigma_{\rm line}$, equal to $c\tau \sigma_{\rm line}^2/G$
(hereafter VP$_{\rm s}$), 
versus the
virial product based on FWHM, equal to $c\tau {\rm FWHM}^2/G$
(hereafter VP$_{\rm f}$), for all the datasets
in Table 1, except for four data sets for which
the lag uncertainty is very large, with 
$\Delta \tau_{\rm cent} > \tau_{\rm cent}$, i.e.,
PG~0844+349, NGC~3227, NGC~4593 and IC~4329A. 
While the virial products based on FWHM are
well-correlated with those based on $\sigma_{\rm line}$,
there is scatter of about a factor of three in each case,
reflecting the broad range of values of 
${\rm FWHM}/\sigma_{\rm line}$ in these datasets
(cf.\ Fig.\ 9 of Peterson et al.\ 2004).

\begin{figure}
\begin{center}
\includegraphics[width=\linewidth]{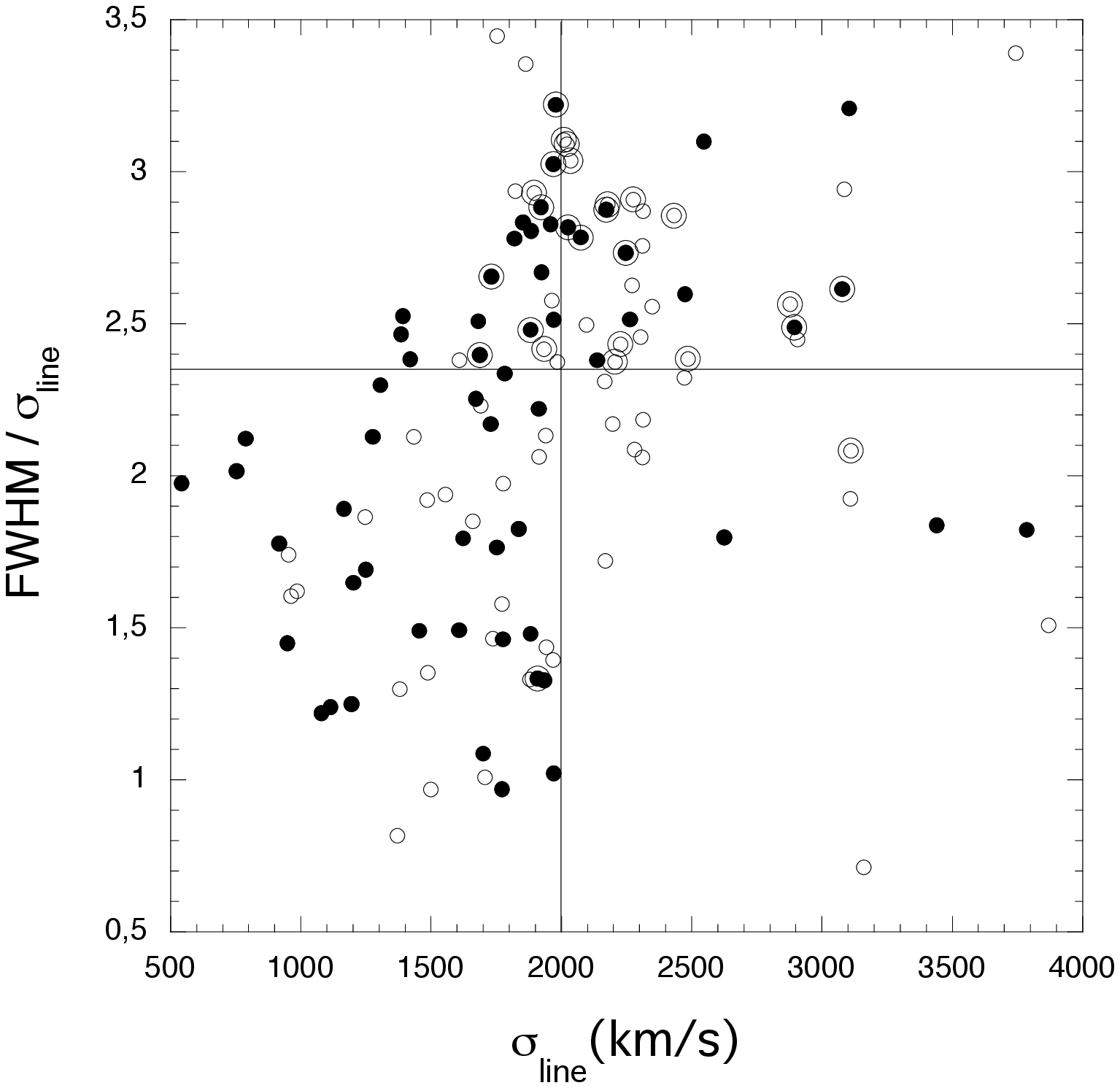}
\caption{ The H$\beta$ line-width ratio
${\rm FWHM}/\sigma_{\rm line}$ versus
$\sigma_{\rm line}$ in all the datasets in Table 1.
The open and filled circles correspond respectively to values based
on mean and rms spectra. The large symbols mark the different
datasets for NGC~5548. The horizontal line is the value of
the ratio for a Gaussian profile, and 
the vertical line is an arbitrary division at
$\sigma_{\rm line} = 2000$\,km\,s$^{-1}$.
The horizontal line divides the samples into our 
Populations 1 (lower) and 2 (upper), and the vertical line approximates
the division of Sulentic et al.\ (2000) into Populations 
A (left)  and B (right).}
\label{fig-FWssig-vs-sig}
\end{center}
\end{figure}

The immediate question now becomes which one of the line-width
measures ought to be used to compute the BH mass? To investigate
this further, we considered how the line-width ratio
${\rm FWHM}/\sigma_{\rm line}$ is correlated with other
spectral properties. In Fig.\ \ref{fig-FWssig-vs-sig},
we show for H$\beta$ ${\rm FWHM}/\sigma_{\rm line}$ as a function of the 
line width $\sigma_{\rm line}$ in both the mean and rms spectra.
The results for the mean and the rms spectrum are quite
similar. In both cases, the ratio has a large dispersion, but 
${\rm FWHM}/\sigma_{\rm line}$ shows
{\it a clear tendency to be smaller than 2.35, the ratio for a
Gaussian profile, for narrow-line objects, and larger than 2.35 for
broad-line objects}. Fig.\ \ref{fig-FWssig-vs-sig} tells us
that the broad-line objects have more flat-topped profiles, 
while the narrow-line objects have more extended wings,
relative to a Gaussian; in other
words, this ratio correlates with the characteristics of the
line profile, which in turn correlate with
other spectral properties as is sometimes embodied in
``Eigenvector 1'' from principal component analysis
(e.g., Boroson \& Green 1992), as is well known. 
We can somewhat arbitrarily separate the AGNs into 
two ``populations'' based on line-width ratio,
Population 1 with ${\rm FWHM}/\sigma_{\rm line} < 2.35$ and 
Population 2 with ${\rm FWHM}/\sigma_{\rm line} > 2.35$.
As seen in Fig.\ \ref{fig-FWssig-vs-sig},
is corresponds very roughly to a division around
$\sigma_{\rm line} = 2000$\,km\,s$^{-1}$, which is approximately
${\rm FWHM} = 4000$\,km\,s$^{-1}$;   
our demarcation is thus interestingly reminiscent of the division of AGNs 
by Sulentic et al.\ (2000)
into a Population A, with ${\rm FWHM} \le 4000$\,km\,s$^{-1}$,
and a Population B, with ${\rm FWHM} \ge 4000$\,km\,s$^{-1}$.
Sulentic et al.\ identify Population A, which 
includes NLS1s, with
low BH mass and high accretion-rate sources, and Population B
with low accretion-rate radio-loud (or pre-/post-cursors of radio-loud)
sources. Apparently the large range of the line-width ratio
${\rm FWHM}/\sigma_{\rm line}$ is telling us
something important about the BLR:
the line-of-sight kinematics and/or the BLR structure show a great variety.
The inclination can also play an important role, as will be discussed later.

\subsection{Effects Other Than Inclination: The Importance of
NGC~5548}

The galaxy NGC~5548 is the best-studied of all 
reverberation-mapped AGNs. 
In Figs.\ \ref{fig-sig-FW-rms-vs-sig-FW-mean} --
\ref{fig-FWssig-vs-sig}, NGC~5548 appears multiple times
as a result of 14 separate years of intensive optical
spectroscopic monitoring first at the Wise Observatory
in 1988 (Netzer et al.\ 1990) then for 13 consecutive
years by the International AGN Watch beginning in
late 1989 (Peterson et al.\ 2002 and
references therein).  We see that all the quantities
in Table 1 vary with time: the FWHM in the mean spectrum and 
$\sigma_{\rm line}$ in both
the mean and the rms spectrum vary by less than a factor of two, but
the FWHM in the rms spectrum varies by up to a factor of four.
Figure \ref{fig-FWssig-vs-sig} shows that 
${\rm FWHM}/\sigma_{\rm line}$ can also vary by a factor larger
than two, and moreover that it can migrate between our two
arbitrarily defined populations, as well as those
of Sulentic et al.\ (2000). This demonstrates clearly
that ${\rm FWHM}/\sigma_{\rm line}$ does not depend solely
on either mass or inclination, as these are constant
over the timescales involved.

\begin{figure*}
\begin{center}
\includegraphics[width=\linewidth]{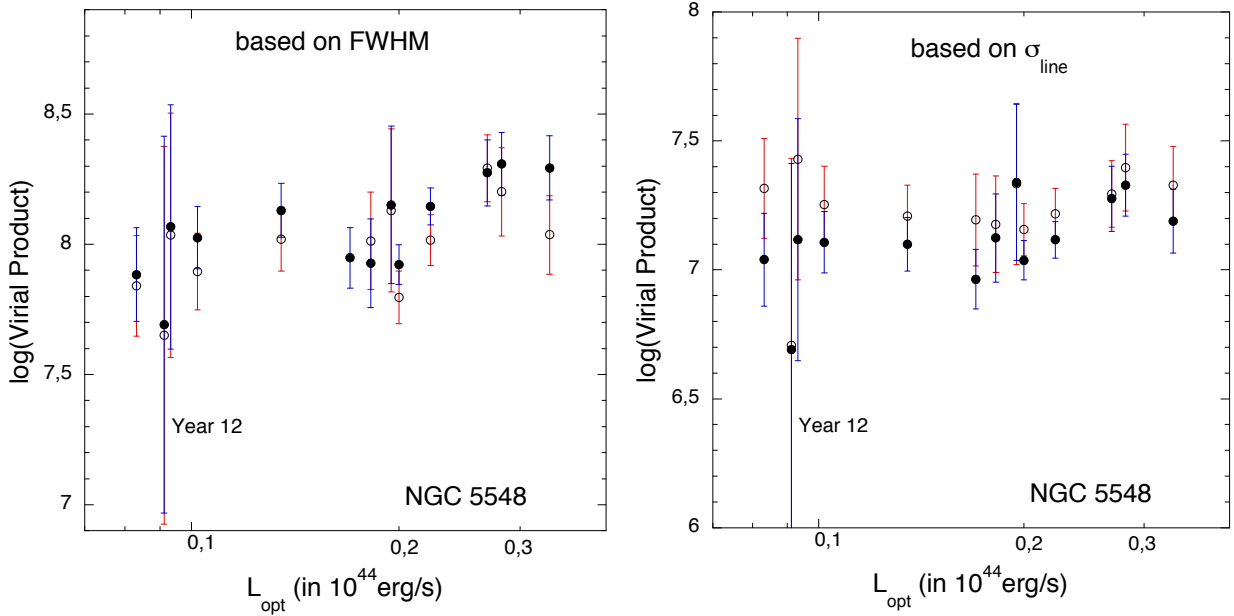}
\caption{The virial product based on the FWHM (left panel) and on $\sigma_{\rm line}$ (right panel), versus the optical luminosity of NGC~5548, 
corrected for the host-galaxy starlight contribution.
The open and filled circles correspond respectively to values
based on the mean and rms spectra. In the left panel, the point corresponding to 
the rms spectrum of Year 5 has been suppressed as 
determination of the FWHM of H$\beta$ is problematic in
this case (see Fig.\ 14 of Peterson et al.\ 2004).}
\label{fig-VP-vs-L-NGC5548}
\end{center}
\end{figure*}

An obvious requirement is that the BH mass, or virial
product, must be constant for all the individual datasets on
NGC~5548. In Fig.\ \ref{fig-VP-vs-L-NGC5548},
we show VP$_{\rm s}$ and VP$_{\rm f}$, as measured in the mean and rms spectra,
as a function of the mean luminosity, for 13 NGC~5548 
datasets\footnote{We have suppressed the point corresponding to the 
Wise Observatory data from 1988 which is
not reliable on account of a variable line-spread function, as noted by 
Peterson et al.\ (2004).}.
In this diagram, the optical continuum luminosity
has been corrected for host-galaxy contamination by using the
value of the starlight contribution
given by Bentz et al.\ (2006).
Despite the large variations in the value of $\sigma_{\rm line}$
as seen in Fig.\ \ref{fig-sig-FW-rms-vs-sig-FW-mean}, 
we see that all the measures of VP$_{\rm s}$ in both
the mean and rms spectra are consistent with a constant value, 
for masses based on either the mean or rms spectra.
The discrepant point is from Year 12 (2000) of the AGN Watch program 
(Peterson et al.\ 2002), which was
the most poorly sampled H$\beta$ light curve in the whole series
and yielded somewhat ambiguous cross-correlation results
(see Fig.\ 2 of Peterson et al.\ 2002).

\subsection{Comparison with Masses Based on $\sigma_*$}

Ferrarese \& Merritt (2000) and Gebhardt et al.\ (2000) 
showed that a tight relationship exists between the BH
mass $M_{\rm BH}$ and the velocity dispersion 
$\sigma_*$ of the stars in the bulge of the host galaxy.  
The galaxies that define the $M_{\rm BH}$--$\sigma_*$ 
relationship are nearly all\footnote{One Seyfert 2 galaxy,
NGC~4258, whose BH mass was measured by megamaser
motions (Miyoshi et al.\ 1995), was included in the sample of galaxies that
defined the $M_{\rm BH}$--$\sigma_*$ relationship.}
quiescent galaxies whose BH masses were 
measured by stellar or gas kinematics.
It is difficult, but possible, to measure bulge
velocity dispersions in at least the lowest-luminosity AGNs.
At the present time, velocity dispersion measurements
have been published for about 16 AGNs, i.e., almost half of the
reverberation-mapped sample. Onken et al.\ (2004) plotted the 
values of VP$_{\rm s}$ versus $\sigma_*$
for these objects and 
obtained a relationship consisent with the 
quiescent-galaxy $M_{\rm BH}$--$\sigma_*$ relationship.
By making the assumption that the AGN
$M_{\rm BH}$--$\sigma_*$ relationship is identical to that
for quiescent galaxies, they were able to convert the
VP$_{\rm s}$--$\sigma_*$ relationship to a
$M_{\rm BH}$--$\sigma_*$ relationship by determining a statistical
value for the scale factor of eq.\ (1). 
Onken et al.\ found  $\langle f \rangle = 5.5 \pm 1.8$.

Here we will carry out a similar exercise, but 
for all four ways of computing the virial product,
i.e. VP$_{\rm s}$ and VP$_{\rm f}$ for both the mean and
rms spectra. We compute for each AGN in the Onken sample
an estimate of the BH mass $M_{{\sigma*}}$ in solar masses from the
formula of Tremaine et al.\ (2002),
\begin{equation}
\log M_{\sigma*} = 8.13  + 4.02 \log 
\left( \sigma_*/200\,{\rm km\,s}^{-1} \right). 
\end{equation}
We have excluded two objects with highly uncertain reverberation-based
masses,  IC 4329A and NGC 4593, leaving a sample of 14 objects
and 39 datasets (herafter the ``Onken sample'').

In Table \ref{table2}, we give the average scale factors, computed
as by Onken et al.\ (2004). We first note that the scale factor
for the entire sample using 
$\sigma_{\rm line}$ measured the rms spectra yields a 
value identical to that obtained by Onken et al.,
$f=5.5$.  We also note that using 
$\sigma_{\rm line}$ measured the {\em mean} spectra yields
$f = 3.85$, reflecting our earlier observation
that on average $\sigma_{\rm line}$ is $\sim20$\% broader in the mean spectra 
than in the rms spectra; the scale factor based on FWHM is 
only 20\% larger for the mean spectra compared to the rms spectra, however,
because FWHM is typically only about 10$\%$ broader in the mean
spectra than in the rms spectra.  We see also that 
${\rm VP_f}/{\rm VP_s} \approx 4$, as expected since $\langle {\rm
FWHM}/\sigma_{\rm line} \rangle \approx 2.$

\begin{table}
\begin{center}
\includegraphics[width=\linewidth]{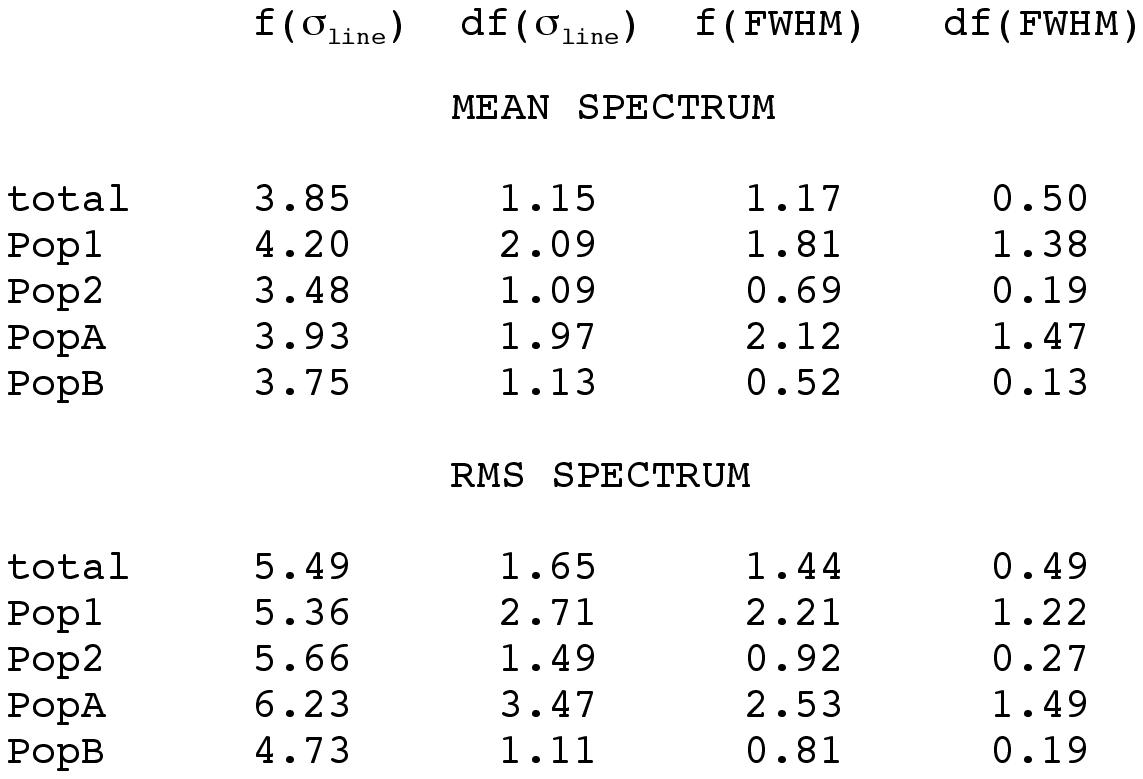}
\caption{The scale factors with their uncertainties for the Onken sample
and for  two populations  
(1) separated at ${\rm FWHM}/\sigma_{\rm line} = 2.35$
(Pop1 and Pop2) as explained in the text and
(2) separated at ${\rm FWHM} = 4000\,{\rm km\,s}^{-1}$
(PopA and PopB) according to Sulentic et al.\ (2000).
 }
\label{table2}
\end{center}
\end{table}

As noted earlier, there is great range in the values of 
${\rm FWHM}/\sigma_{\rm line}$ and 
thus the scale factors that we would compute will
be strongly dependent on the value of this ratio
for the objects in the sample. This leads us to compute the
scale factors separately, again for all four ways of computing
the virial product, for our two populations,
Population 1 with ${\rm FWHM}/\sigma_{\rm line} < 2.35$ and 
Population 2 with ${\rm FWHM}/\sigma_{\rm line} > 2.35$.
These values are also given in Table \ref{table2}.
We have also computed separate scale factors for
Population A (${\rm FWHM} < 4000\,{\rm km\,s}^{-1}$) and 
Population B (${\rm FWHM} > 4000\,{\rm km\,s}^{-1}$) of
Sulentic et al. (2000).
There are 9 objects in the Onken sample in Population 1, among which the
6 objects of Population A are all included. The 6 common objects
are NGC 4051, NGC 3783, NGC 7469, Mrk 110, Mrk 590, and
3C 120. Population B consists of 8 objects, among which the 5 objects
of Population 2 are all included, namely, NGC 4151, NGC 5548, Mrk 817, 
Akn 120, and 3C 390.3.

The results of this exercise are very revealing:
there is a clear difference between the scale factors of the different
subsets, especially if we compare the objects common to Populations 1
and A with those common to Populations 2 and B. 
For VP$_{\rm f}$, Populations 1 and A have
{\it scale factors larger by a factor of
$\sim3$} than those for Populations 2 and B, both in the mean
and rms spectra. This trend does not exist for
${\rm VP_s}$; in both the mean and rms spectra, the scaling
factors computed from ${\rm VP_s}$ are consistent with
a constant value. 
It is clear that the statistics are poor, and that more objects in the
sample are urgently needed. Nevertheless some conclusions can already
be drawn from these results:

\begin{enumerate}

\item Since scale factors based on $\sigma_{\rm line}$ for different
populations are consistent with a single value, contrary to the case
for scale factors based on FWHM, ${\rm VP_s}$ is not sensitive to
the line profile, the Eddington ratios, inclination, or whatever
factors distinguish between Populations 1 and 2 or A and B, 
whereas ${\rm VP_f}$ clearly is affected by these factors.
In other words, {\it $\sigma_{\rm line}$ is a less biased measure of 
the velocity dispersion than is} FWHM.

\item Use of FWHM as line-width measure and applying a
constant scale factor corresponding to the average for the whole
sample {\it will underestimate the masses of Populations 1 and
A} (including NLS1s) and {\it overestimate the masses of 
Populations 2 and B}. This could be a contributing factor to why
high ratios of bulge-to-BH masses have been
found in NLS1s (Wandel 2001; Mathur et al.\ 2001), since the BH
masses are based on FWHM in these works. On the other
hand, since it appears that the FWHM is influenced by some still
undefined physical parameters to which $\sigma_{\rm line}$ is not
sensitive, it could be used as an indicator of 
whatever these currently unidentified parameters turn out to be.
We argue below that one unidentified parameter is BLR inclination, 
and another is the Eddington ratio.

\item {\it It is possible to correct ${\rm VP_f}$ to obtain an
unbiased BH mass.} At the present time, the correction is only
approximate because the uncertainties on the scale factors are
large. {According to Table 2, a simple prescription is to use $f =
2.35$ for Population 1 and $f = 0.85$ for Population 2 for the rms
spectrum, and $f = 1.5$ for Population 1 and $f = 0.50$ for Population
2 for the mean spectrum, so the mass\footnote{It is perhaps interesting that prior to the actual
calibration of the reverberation-based mass scale by Onken et al.\
(2004), it was common to assume $f=0.75$ (Netzer 1990; Wandel et al.\
1999; Kaspi et al.\ 2000) for FWHM-based mass estimates.} is given by $M_{\rm BH} = f {\rm
VP_f}$}.  But to
avoid a discontinuity between these two rather arbitrarily defined
populations, we tentatively suggest using
for the rms spectrum,
\begin{eqnarray}
\label{eq:FWHMscalefactor1}
f & = & 2.35\ \ \ {\rm (for\ (FWHM}/\sigma_{\rm line}) \le 1.4) \\
  & = & 3.85 - 1.07 {\rm (FWHM}/\sigma_{\rm line})\ \nonumber \\
  &   & \ \ \ \ \ (\rm for\ 1.4 < {\rm (FWHM}/\sigma_{\rm line}) < 2.8)
\nonumber\\
  & = & 0.85\ \ \ {\rm (for\ (FWHM}/\sigma_{\rm line}) \ge 2.8),
\nonumber
\end{eqnarray}
and for the mean spectrum,
\begin{eqnarray}
\label{eq:FWHMscalefactor2}
f & = & 1.5\ \ \ {\rm (for\ (FWHM}/\sigma_{\rm line}) \le 1.4) \\
  & = & 2.5 - 0.71 {\rm (FWHM}/\sigma_{\rm line})\ \nonumber \\
  &   & \ \ \ \ \ (\rm for\ 1.4 < {\rm (FWHM}/\sigma_{\rm line}) < 2.8)
\nonumber\\
  & = & 0.5\ \ \ {\rm (for\ (FWHM}/\sigma_{\rm line}) \ge 2.8).
\nonumber
\end{eqnarray}

It is rarely the case, however, that $\sigma_{\rm line}$ measurements
are available, especially in published compilations of line widths.
in such cases, we suggest using for the rms spectrum,
\begin{eqnarray}
\label{eq:FWHMscalefactor3}
f & = & 2.35\ \ \ {\rm (for\ (FWHM} \le 2000\,{\rm km\,s}^{-1}) \\
  & = & 3.1 - 1.5\ {\rm (FWHM}/4000\,{\rm km\,s}^{-1}) \nonumber \\ 
  &   & \ \ \ \ \ (\rm for\ 2000 < {\rm FWHM} < 6000\,{\rm km\,s}^{-1})
\nonumber\\
  & = & 0.85\ \ \ {\rm (for\ FWHM} \ge 6000\,{\rm km\,s}^{-1}),
\nonumber
\end{eqnarray}
and for the mean spectrum,
\begin{eqnarray}
\label{eq:FWHMscalefactor4}
f & = & 1.5\ \ \ {\rm (for\ (FWHM} \le 2000\,{\rm km\,s}^{-1}) \\
  & = & 2 - {\rm (FWHM}/4000\,{\rm km\,s}^{-1}) \nonumber \\ 
  &   & \ \ \ \ \ (\rm for\ 2000 < {\rm FWHM} < 6000\,{\rm km\,s}^{-1})
\nonumber\\
  & = & 0.5\ \ \ {\rm (for\ FWHM} \ge 6000\,{\rm km\,s}^{-1}). 
\nonumber
\end{eqnarray}

\end{enumerate}

For $\sigma_{\rm line}$-based data, for all AGNs we suggest using
the value $f = 5.5$, (i.e., $M_{\rm BH} = 5.5 {\rm VP_s}$) for the rms
spectrum, as proposed by Onken et al. (2004), and $f = 3.85$ (i.e.
$M_{\rm BH} = 3.85 {\rm VP_s}$) for the mean spectrum.

\subsection{Application of the Scale Factors: A Consistency Test}

In the last section, we provided 
scaling factors to convert both ${\rm VP_s}$
and ${\rm VP_f}$ into BH masses. We now apply these to the
entire sample of 35 objects from Table 1.
	
\begin{figure*}
\begin{center}
\includegraphics[width=\linewidth]{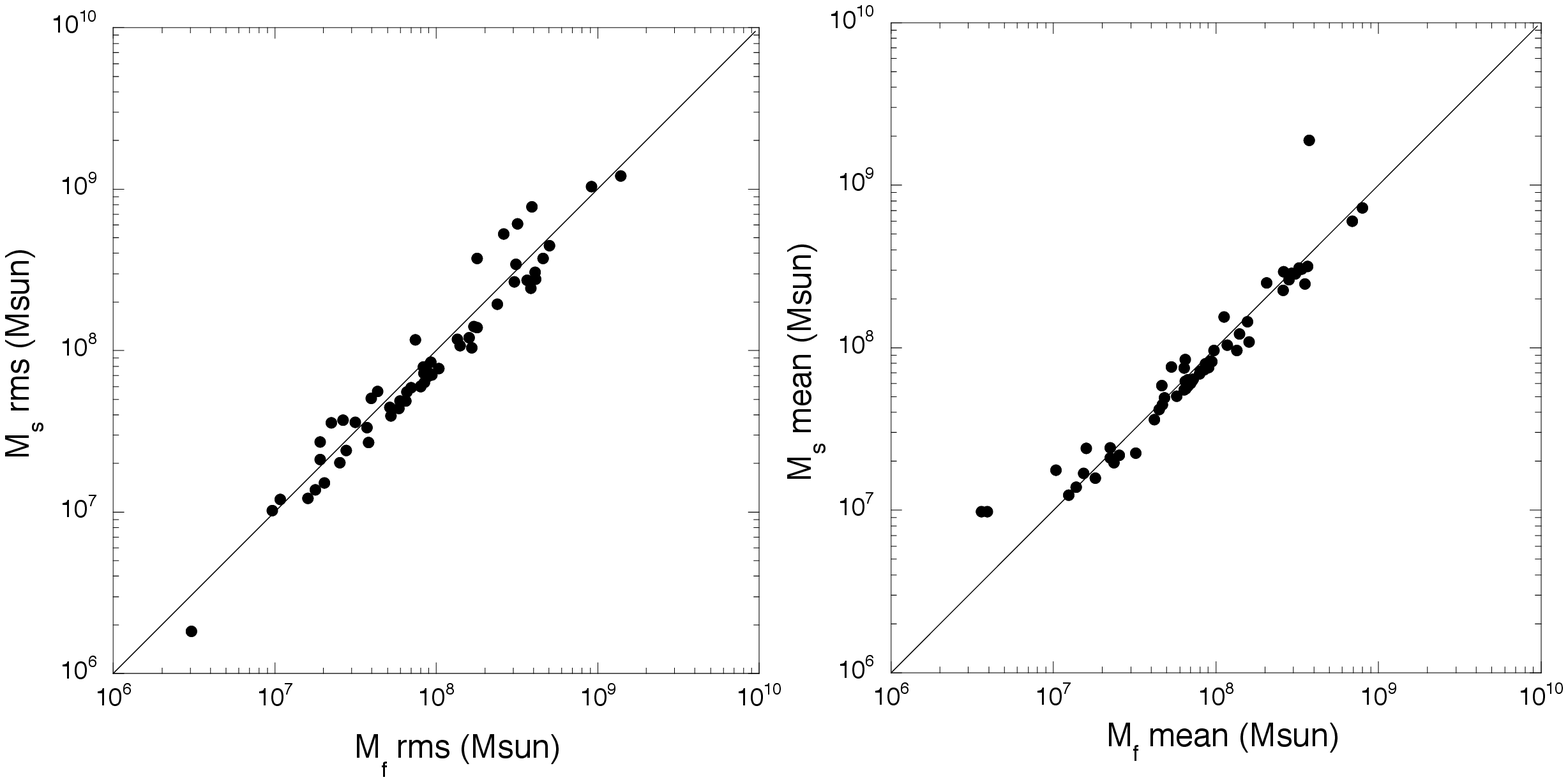}
\includegraphics[width=\linewidth]{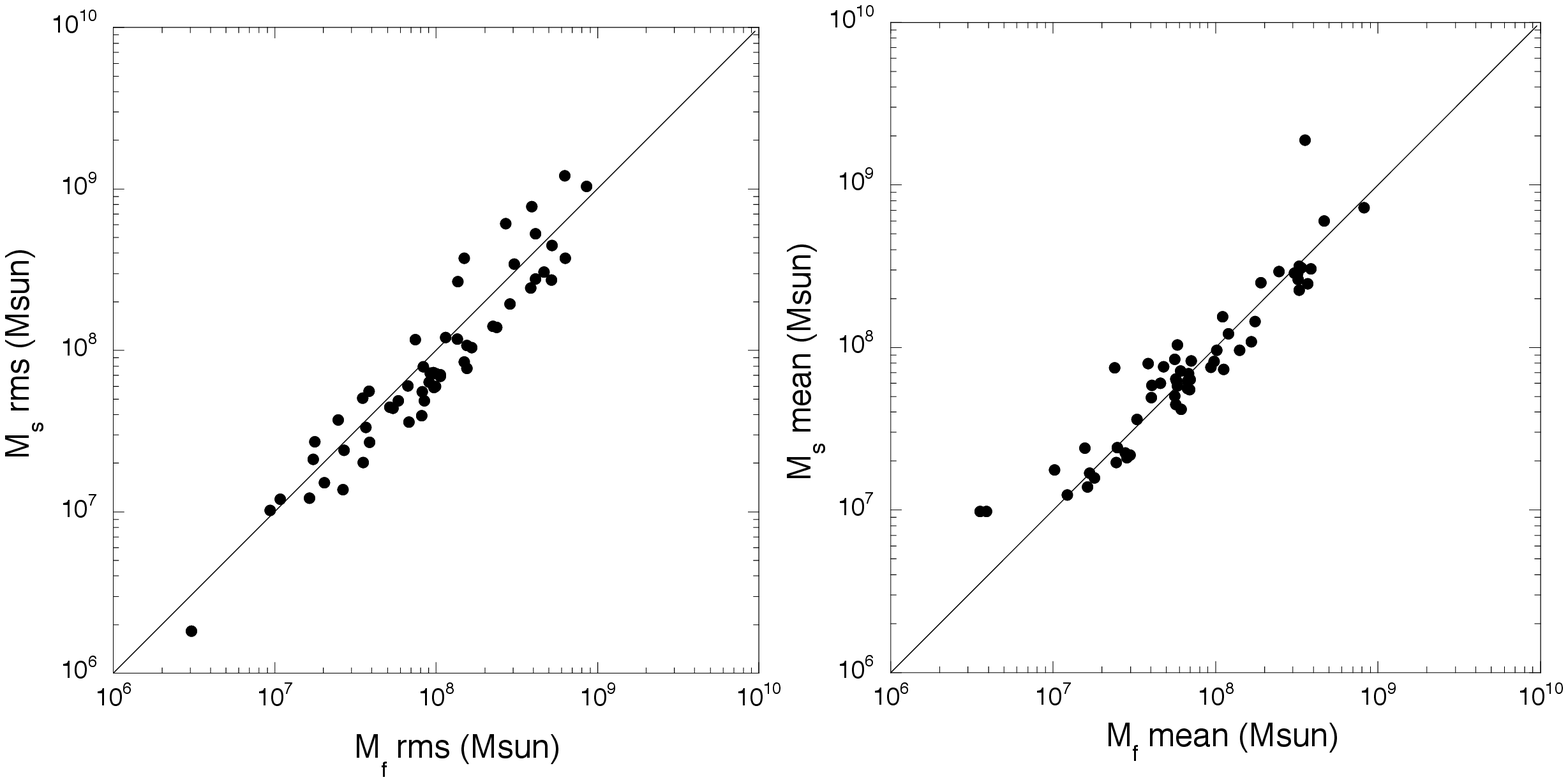}
\caption{Mass estimates based on  $\sigma_{\rm line}$
as a line-width measure are plotted versus masses based
on FWHM measurements and adjusted according to 
eqs.\ (\ref{eq:FWHMscalefactor1}) and (\ref{eq:FWHMscalefactor2}) 
(upper panels) and to eqs.\ (\ref{eq:FWHMscalefactor3}) and 
(\ref{eq:FWHMscalefactor4}) (lower panels). The data in the left panels
are based on line widths in the rms spectrum 
and those based on the mean spectrum are shown in the
right panels. The same four datasets as in Fig.\ \ref{fig-VPsig-vs-VPFW}
have been suppressed. The straight lines trace the diagonals.}
 \label{fig-Msig-vs-MFW}
 \end{center}
\end{figure*}

Figure \ref{fig-Msig-vs-MFW} compares the $\sigma_{\rm line}$-based
masses, computed using $f= 5.5$ and $f=3.85$ for the
rms and mean spectra, respectively, 
with the corrected FWHM-based masses in both the mean
and rms spectrum for all the data sets in the sample. In the upper
panels, the FWHM-based masses are 
corrected by using eqs.\ (\ref{eq:FWHMscalefactor1})
and (\ref{eq:FWHMscalefactor2}), while in the lower panels, the
corrections are given by eqs.\ (\ref{eq:FWHMscalefactor3}) and 
(\ref{eq:FWHMscalefactor4}).  The two formulations give very similar
results, and the scatter is reduced compared to Fig.\
\ref{fig-VPsig-vs-VPFW}, even when only the FWHM used for the
correction. This demonstrates the utility of using eqs.\
(\ref{eq:FWHMscalefactor3}) and (\ref{eq:FWHMscalefactor4}) to adjust
FWHM-based mass estimates when $\sigma_{\rm line}$ is not known.
		
\subsection{Line Width Measures and the Eddington Ratio}

\begin{figure}
\begin{center}
\includegraphics[width=\linewidth]{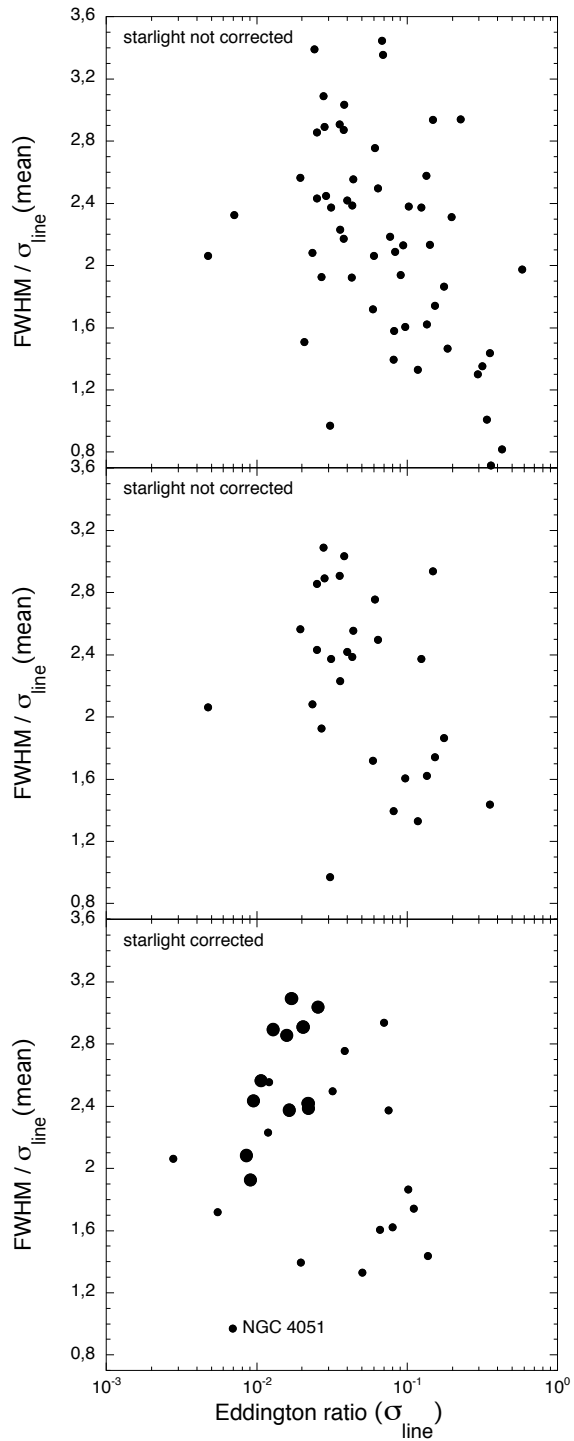}
\caption{The line-width ratio ${\rm FWHM}/\sigma_{\rm line}$ 
for H$\beta$ in the mean spectra,
versus the Eddington ratio (based on masses from Peterson et al.\ 2004)
for all the datasets for the 35
reverberation-mapped objects in Table 1 (top panel).
 The middle panel shows the subset of points from the top panel
which can be corrected for the host-galaxy starlight contribution
to the luminosity using values from Bentz et al.\ (2006).
The bottom panel shows the points from the middle panel after
correction for starlight.} The larger points in the lower
panel represent independent data sets on NGC 5548.
Assumptions used in computing the Eddington
ratio are described in the text. 
\label{fig-FWssig-vs-Redsig}
\end{center}
\end{figure}

The line-width ratio ${\rm FWHM}/\sigma_{\rm line}$ is 
of potential importance as it may trace physical parameters
in the inner regions of AGNs. Since we have already pointed
out a qualitative correlation with Eigenvector 1 properties,
which has been argued to measure the Eddington ratio (e.g., Boroson 2002), it is
now of interest to look at this more quantitatively.
In Fig.\ \ref{fig-FWssig-vs-Redsig}, we show the line-width
ratio ${\rm FWHM}/\sigma_{\rm line}$ as measured in the
mean spectrum (since it is less noisy than the rms spectrum)
as a function of
Eddington ratio, which has been computed assuming that
the bolometric luminosity is $L_{\rm bol} \approx 10 L_{\rm opt}$ 
(based loosely on Kaspi et al.\ 2000 and Elvis et al.\ 1994), 
and $M_{\rm BH}$ is based on measurements of $\sigma_{\rm line}$
in the rms spectrum, which appears to yield the most
accurate mass estimate (Peterson et al.\ 2004).
The top panel of Fig.\ \ref{fig-FWssig-vs-Redsig} shows a clear anticorrelation between
the line-width ratio ${\rm FWHM}/\sigma_{\rm line}$ and
the Eddington rate, though with considerable
scatter.
The dependency of ${\rm FWHM}/\sigma_{\rm line}$ on
Eddington rate is physically quite plausible as
we expect that 
that the structure and the dynamics of the BLR, which determine
the line profile, depend on the
accretion rate.
 We note, however, that the Eddington rates in this figure are somewhat overestimated
because the optical luminosities have not been corrected for
contamination by host-galaxy starlight. At the present time,
it is possible to accurately correct for the host-galaxy
contribution to the luminosity for only a subset of these
AGNs, those observed by Bentz et al.\ (2006). In the middle panel
of Fig.\ \ref{fig-FWssig-vs-Redsig}, we show the subset of points
from the top panel for which Bentz et al.\ provide measurements of
the starlight contribution. In the bottom panel, we show the
points from the middle panel after correction for starlight.
The anticorrelation that is clearly seen in the top panel appears
to have vanished in the middle and bottom panels. This is simply
because most the high-Eddington rate objects are PG quasars
which were not observed by Bentz et al. Given the higher luminosities
of these sources, the starlight corrections are likely to be small
so the points in the lower right of the top panel will have very
nearly the same positions in the bottom panel, thus preserving
the anticorrelation.
In the lower panel, we highlight by use of larger symbols the
multiple independent measurements of NGC 5548. These are dispersed
in a direction almost normal to the anticorrelation seen in the
top panel, which suggests that much of the scatter in this
relationship can be attributed to intrinsic variability.
We will discuss this further in a future paper.
		
To summarize this section, we have shown that we can crudely separate
AGNs into high Eddington ratio objects
whose spectra are characterized by small values of 
${\rm FWHM}/\sigma_{\rm line}$ (Population 1) and low Eddington ratio objects, which
have large values of 
${\rm FWHM}/\sigma_{\rm line}$ (Population 2).
 We find that $\sigma_{\rm line}$ is an unbiased estimator of the BH mass,
   whereas FWHM requires an adjustment for its sensitivity to still
   undefined physical parameters, likely to be the Eddington ratio
   and/or the source inclination. Finally, we note that 
${\rm FWHM}/\sigma_{\rm line}$  is not correlated with BH mass, or
with luminosity.
	
\section{Influence of the BLR Inclination}

Since we observe only the line-of-sight velocity distribution of the
BLR, it is obvious that inclination of the BLR will play a significant
role in the value of the scale factor except in the unlikely case of
an isotropic velocity field. For instance, Krolik (2001) pointed
out that if the BLR is a thin disk, it would lead to an
underestimation of the mass by one or two orders of magnitude for
objects seen at low inclination  To study the influence of the
inclination, it is necessary to have an idea of the structure and the
dynamics of the BLR. In this section, we show that a few simple but
plausible structures can be parameterized in a common fashion and we
explore the effects of inclination in this context as an exercise.

\subsection{Structure and Dynamics of a Rotationally Dominated BLR}

It is often noted that the structure and the dynamics of the BLR are
not known, despite more than thirty years of intense studies. Indeed
we only know with certainty, thanks to reverberation studies, that the
region emitting the ``low-ionization lines'' like H$\beta$ is
gravitationally bound to the BH (Peterson \& Wandel 1999, 2000;
Onken \& Peterson 2002; Kollatschny 2003a) and more
precisely that the bulk radial velocity is small 
(Clavel 1991). But it remains true that 
we do not know if the BLR is a spherical structure dominated by
turbulent motions, if it includes a wind component, or if it is a disk
dominated by rotational motions,

There is actually fairly strong evidence for the latter case in radio-loud
AGNs. Based on the ratio of the radio core flux to the extended
radio lobe flux, $R$, which is related to the relativistic amplification
of the core source and is large when the object is seen face-on, 
Wills \& Browne (1986) and Jackson \& Browne (1991) found 
a lack of broad lines for face-on objects. 
Also, Vestergaard, Wilkes, \& Barthel (2000) find that the width
of the base of the \ion{C}{IV}\,$\lambda1549$ emission line is
inversely correlated with $R$, suggesting the 
existence of a largely radial disk wind. In
superluminal objects where the inclination can be derived quite
precisely, Rokaki et al.\ (2003) have shown that the line width is
anti-correlated with several beaming indicators, and is consistent
with a disk structure of the BLR. Finally, double-peaked profiles of
the Balmer lines, though observed only in a small fraction of radio
AGNs, are another suggestion of disk structure (Eracleous \& Halpern
1994 and subsequent papers). Such double peaks, more or less pronounced, are
characteristics of our Population 2 AGNs.

There is no direct evidence that this picture also applies to 
radio-quiet objects, but several observations fit this model quite
well (e.g., see Smith et al.\ 2005). There is also {\it indirect}
evidence which cannot be ignored. A common view of the BLR is
that it consists of 
a large number of clouds with high velocities surrounding a
central source of UV--X-ray radiation that photoionizes the clouds and
produces line emission by reprocessing. These clouds should have a
large covering factor, since at least 10\% of the central X-UV
source, and in some cases a larger fraction, has to be absorbed by
the BLR in order to account for the large measured equivalent widths
of the broad lines. On the other hand, the column density of the
clouds is inferred to be high, $10^{23-24}\,{\rm cm}^{-2}$. There is actually no
observational limit to the real column density of the clouds, as a
large fraction can stay neutral and unobservable (for a review, see
Collin-Souffrin \& 
Lasota 1988, for example). The lack of Lyman edges in absorption
and of damped Ly$\alpha$ lines in AGN spectra is difficult to explain
in this context, unless the BLR clouds are not located on the line of
sight to the central source. Such a configuration is natural in the
framework of unified schemes (Antonnucci \& Miller 1985), if the
BLR has a disk structure in the same plane as that of the obscuring
``torus" and of the inner disk. Since for Seyfert 1 galaxies the central continuum is seen from above at relatively
   small inclinations to the disk normal, it would not be absorbed by
   this BLR disk.

As explained above, the BLR disk must ``see'' the central source, as it
must be able to capture a large fraction of its ionizing photons. 
We can thus
immediately eliminate the possibility that the BLR disk is a thin,
flat structure that is directly illuminated by the central UV--X-ray source.
In this scenario,
the central source should have a large scale height above the disk, of the
order the radius of the BLR disk itself. But we know from the study of the continuum
emission, in particular from the correlations between the UV and X-ray flux
variability, that the UV--X-ray source is located much closer to the BH
than the BLR, typically at distances of 
10 to $10^2\,R_{\rm G}$, where $R_{\rm G}$ is 
the gravitational radius, while the distance of the BLR is of the
order of $10^3$ to $10^4\,R_{\rm G}$. 
A possible exception could be the case of low-luminosity AGNs with very
broad double-peaked profiles, where it has bee suggested that
the inner part of the disk might be
an inflated ion-supported torus (cf.\ \S{\ref{section:discussion}}).

Another possibility is that the surface of the BLR disk is illuminated
indirectly by the central source, which could occur
if its radiation is backscattered by
a hot medium (or corona) located above the BLR disk, as proposed by Dumont \&
Collin-Souffrin (1990). This model faces some difficulties, however, 
as the hot
medium should have a Thomson thickness at least of the order of
unity. In this case, the variations of the central
source would be smoothed by multiple electron scattering, and
the shape of the observed X-ray continuum would be modified by
absorption and Compton diffusion (Reynolds \& Wilms 2000), two predictions
that are contradicted by the observations. However the recent observations of
very thick X-ray winds in some NLS1s (Pounds et al.\ 2003) and the
suggestion by Gierlinski \& Done (2004) and Chevallier et al.\ (2005)
that the soft X-ray excess is due to the absorption by a very thick
warm absorber, could perhaps rehabilitate this idea for high
accretion-rate objects. But it cannot be considered as a 
general solution for the BLR of all AGNs.

Thus it is unlikely that the BLR disk is a very thin flat
structure entirely dominated by rotation, and other possibilities
must be considered\footnote{It is worthwhile recalling that when we
speak of a ``disk,'' it does not necessarily mean a continuous medium, 
but could refer to a clumpy medium made of an ensemble of clouds.}.

\subsubsection{The BLR as a Flared Disk}

The required illumination of the BLR implies that the opening angle of
the BLR disk should be large, i.e., $\Omega/4\pi \ge 0.1$.  In other words,
the BLR disk should have an aspect ratio 
larger than $H/R=0.1$ at the
location of the line-emitting region, where $H$ is the disk thickness
at the radius $R$. Moreover it should be ``flared,'' i.e.,
that $H$ increases more rapidly than linearly with increasing $R$. 

Such flared disks are predicted in the context of the ``standard
model'' (Shakura \& Sunyaev 1973), and were invoked by 
Dumont \& Collin-Souffrin
(1990) as being the origin of the low-ionization lines in AGNs.  In
this case, the BLR disk should be sustained vertically by a pressure
which is most probably provided by a turbulent velocity of the order
$V_{\rm Kep}H/R$, where $V_{\rm Kep}$ is the local Keplerian
velocity at the distance $R$.  In this model, the {\it observed} value
of $\Delta V_{\rm obs}$ is given by
\begin{equation}
\Delta V_{\rm obs} \approx   V_{\rm Kep}
\left[ (H/R)^2+ \sin^2 i \right]^{1/2},
\label{eq-thickdisk}
\end{equation}
where $i$ is the inclination of the system to our line of sight.

\subsubsection{The BLR as a Warped Disk}
A configuration which could be invoked for effective
illumination of the disk by the
central source is a warped disk structure. 
Such a structure is observed in the case of 
NGC~4258 through the maser spots (Greenhill et al.\ 1995; Miyoshi et
al.\ 1995). In this object, the rotation law at the distance 
of the megamaser sources is Keplerian, but the disk is tipped
downward by about $10^{\rm o}$, which allows more of the disk
surface to be directly exposed to the central source.
Wijers \& Pringle (1999) have suggested that similar warping
should arise in AGNs as a response of the disk to the radiation force
from the central source, which can cause the disk to tilt out of the
orbital plane and to precess. However, NGC~4258 is a low-luminosity
AGN, and it is not clear that this would be the case in more luminous objects,
which in particular would be more sensitive to the gravitational
instability.

In this model, the velocity is quasi-Keplerian at any radius, so
$\Delta V_{\rm obs}$ is expressed as
\begin{equation}
\Delta V_{\rm obs} \approx  V_{\rm Kep} \sin i,
\label{eq-warp}
\end{equation}
but the inclination $i$ varies with the radius, i.e., with the 
particular line under
consideration. Another difference from a flat thin disk is that the 
line-emitting region extends over a range of radii, therefore 
over a range of inclinations. There is thus
a lower limit of the inclination, $\Delta i$, 
while it can be arbitrarily small for a flat disk.

\subsubsection{A Two-Component BLR: A Disk and a Wind}

In recent years, arguments supporting the
presence of disk winds have won some popularity on account of
the ability of such structures to explain a number of observed phenomena
such as X-ray and UV absorption, line emission, reverberation results, and
some differences among Seyfert galaxies, quasars, 
broad-line radio galaxies,
and the presence or absence of double-peaked emission-line profiles.
The importance of outflows that are commonly seen in AGNs as 
``warm absorbers'' in X-rays have been recognized through
observations made with {\em XMM-Newton} and
{\em Chandra X-Ray Observatory}. 
It has been suspected for a long time that the BLR has two
components, one that is disk-like and other that is some kind of
outflow, probably a disk wind, either magnetohydrodynamically
or radiatively driven.  The wind scenario also obviates
the problem of having to confine the line-emitting clouds.
Key papers on the topic of disk
winds are by Murray \& Chiang (1995, 1998) and by Proga \& Kallman (2004). In
this model, the broad emission lines are emitted from the base of the
disk winds (see also Elvis 2000).

A plausible configuration is thus a BLR made of two dynamically
distinct components, a disk and a wind. The velocity dispersion 
could be written as 
\begin{equation}
\Delta V_{\rm obs} \approx   
\left[
\alpha^2 \left( a^2 + \sin^2 i \right) V_{\rm Kep}^2 + 
\beta^2  V_{\rm out}^2 \cos^2 i  \right]^{1/2},
\label{eq-deux-composantes}
\end{equation}
where $V_{\rm out}$ is the outflow velocity, assumed to be normal
to the disk, and $\alpha$ and $\beta$ are the contributions of the
thick disk and of the wind, respectively.

However, the $V_{\rm out} \cos i$ term in
eq.\ (\ref{eq-deux-composantes}) cannot typically dominate the H$\beta$ line,
or the H$\beta$ wing would display a strong asymmetry due to
absorption on the receding part of the wind located on the far side of the
accretion disk relative to the central source as seen by the observer. This is however observed in the most extreme objects of
Population A which also show an extended blue wing in the 
high-ionization line \ion{C}{IV}\,$\lambda1549$
(cf.\ Sulentic et al.\ 2000), or in NLS1s which
also show a blue wing in the [\ion{O}{III}]\,$\lambda\lambda4959,$ 5007 lines 
(cf.\ Zamanov et al.\ 2002;
Aoki, Kawaguchi, \& Ohta 2005; Boroson 2003). So, with the exception
of these very
high Eddington ratio objects (say, with $L_{\rm bol}/L_{\rm Edd} \ge 0.3$), of
which we have none in our Population 1 sample, one can neglect the
second term in eq.\ (\ref{eq-deux-composantes}).

\subsubsection{A Simple Parameterization}

We see that, to a zeroth approximation, in all these cases
$\Delta V_{\rm obs}$ can be represented by the expression,
\begin{equation}
\Delta V_{\rm obs} \approx  \left( a^2 + \sin^2 i\right)^{1/2} V_{\rm Kep},
\label{eq-general}
\end{equation}
where the parameter $a$ can be identified with $H/R$ or with $V_{\rm
turbulent}/V_{\rm Kep}$, and it can take a null value in the case of a
warped disk. 

If we identify $V_{\rm Kep}$ with the virial velocity, the virial
product in this parameterization, which we will call the 
``generalized thick disk,'' or more simply
``thick disk,'' even though we
have seen that it can describe other structures, will be
given by
\begin{equation}
{\rm VP}_{\rm thick\ disk}= {R_{\rm BLR}\ V_{\rm Kep}^2\over G}\ 
= {R_{\rm BLR} V_{\rm obs}^2\over G(a^2+\sin^2 i)}.
\label{eq-VP-disque}
\end{equation}

\subsection{Theoretical Considerations}

The ratio of the virial product in the thick disk model 
${\rm VP_{\rm thick\ disk}}$
to the virial product in the general case, 
which we will call the ``isotropic case,''
simply because it does not have cylindrical symmetry, is equal to
\begin{equation}
A={{\rm VP} _{\rm thick\ disk}\over {\rm VP} _{\rm isotropic}}
= {1\over a^2+\sin^2 i}.
\label{eq-G}
\end{equation}

\begin{figure}
\begin{center}
\includegraphics[width=\linewidth]{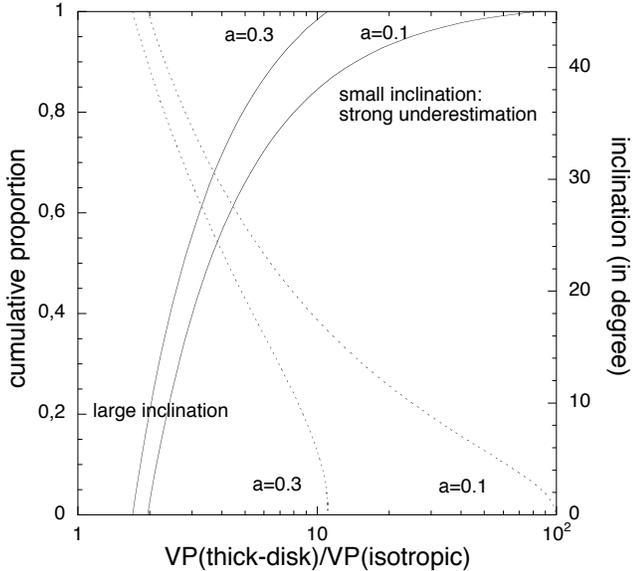}
\caption{The solid curves show the cumulative fraction (left axis)
of AGNs that will have 
$A = {\rm VP}_{\rm thick\ disk}/{\rm VP}_{\rm isotropic}$ smaller
than the value on the bottom axis
for two values of the parameter $a$, assuming that the maximum
inclination is $45^\circ$.
The dotted lines show the value of $A$ for values of the
inclination (right axis) over the range
$0^\circ \leq i \leq 45^\circ$.}
\label{fig-cum-VPdisk-on-VPisotrop}
\end{center}
\end{figure}

The value of the inclination is given as a function of $A$ in
Fig.\ \ref{fig-cum-VPdisk-on-VPisotrop}.  We see that if the thick
disk model is correct, the virial product computed in the general
case could be strongly underestimated for small inclinations. The
general relation can even lead to an underestimate of the mass by
a factor of 100 for $a=0.1$.  Note that we have neglected
any correction due to possible anisotropy of the optical flux
(for instance, the flux emerging from the disk can
be subject to ``limb-darkening'' effects),
or to any obscuration of the BLR. These effects
should be small compared to those we are looking for.

We can compute how many objects would have a given value of $A$ by
assuming that they are distributed at random for inclinations $i \le
i_{0}$. According to the unified scheme (Antonucci \& Miller 1985),
Seyfert 1 nuclei are not seen edge on, and one can quite reasonably
assume $i_{0}\approx 45^{\circ}$, though we have checked that our
conclusions are not qualitatively changed for $ i_0 = 60^{\circ}$.  
The probability of seeing an object at an inclination
angle $i$ per unit angle interval is $\sin i/ (1 - \cos i_{0})$.  The
number of objects per unit interval of $A$ is thus
\begin{equation}
{dN\over dA}= {1\over 2(1 - \cos i_0)}
{ (a^2+\sin^2 i)^{2}\over {\cos i}}.
\label{eq-dNdG}
\end{equation}
Figure \ref{fig-cum-VPdisk-on-VPisotrop} shows the 
integral of this
expression, i.e., the cumulative fraction of AGNs
for which ${\rm VP}_{\rm thick\ disk}/{\rm VP}_{\rm
isotropic}$ is smaller than a given value.  For reasonable values,
say $0.1 \la a \la 0.3$, 
VP could underestimate the mass by as much as one order 
of magnitude in a few to several percent of the objects, 
specifically  those which are seen at small
inclinations. It is thus not implausible that a 
significant fraction of NLS1s, which constitute only 
$\sim10$\% of the local AGN population, 
are actually seen almost ``face-on,'' and
that their Eddington ratio is consequently strongly 
overestimated. We elaborate on this below.

\subsection{Comparison with Observations}

Some authors (Wu \& Han 2001; Zhang \& Wu 2002; McLure 
\& Dunlop 2001) assume that the true BH masses in AGNs satisfy the 
$M_{\rm BH}$--$\sigma_*$ relationship and that the 
differences between the masses deduced from the stellar 
velocity dispersion $M_{\sigma*}$ and the reverberation 
masses $M_{\rm rev}$ are due only to inclination 
effects. They then
proceed to deduce the inclinations for individual AGNs, 
based on the discrepancies between $M_{\rm rev}$ and 
$M_{\sigma*}$.  Wu \& Han and Zhang \& Wu studied 
the reverberation-mapped objects and assumed that the 
BLR was a thin disk with no isotropic component of the 
velocity. McLure \& Dunlop studied a sample of 30 
quasars where the BH masses were estimated using FWHM 
and the $R$--$L$ relationship of Kaspi et al.\ 
(2000) and adopted an ad hoc complex (and rather 
implausible) disk geometry. 

We do not think that it is possible to deduce 
individual inclinations in this way. It would be possible
only if there were no intrinsic scatter either in 
the $M_{\rm BH}$--$\sigma_*$ relationship or in
$M_{\rm rev}$; we know from the best-studied AGN, 
NGC~5548, that VP$_{\rm s}$ and thus $M_{\rm rev}$ has
an intrinsic scatter of about a factor of three.

Here we take a somewhat different approach with the 
goal of testing in a statistical fashion whether 
inclination effects might plausibly account for the 
distribution of differences between $M_{\rm rev}$ and 
$M_{\sigma*}$. We compare the observed distribution of
$M_{\sigma*}/{\rm VP}$ with the theoretical distribution 
${\rm VP}_{\rm thick\ disk}/{\rm VP}_{\rm isotropic}$ 
of
Fig.\ \ref{fig-cum-VPdisk-on-VPisotrop}, since these
should be identical to within a scaling constant (that 
converts VP to $M_{\rm rev}$ at $i = 0^\circ$) if inclination alone is 
responsible for the discrepancies between $M_{\rm rev}$ 
and $M_{\sigma*}$.

\begin{figure*}
\begin{center}
\includegraphics[width=\linewidth]{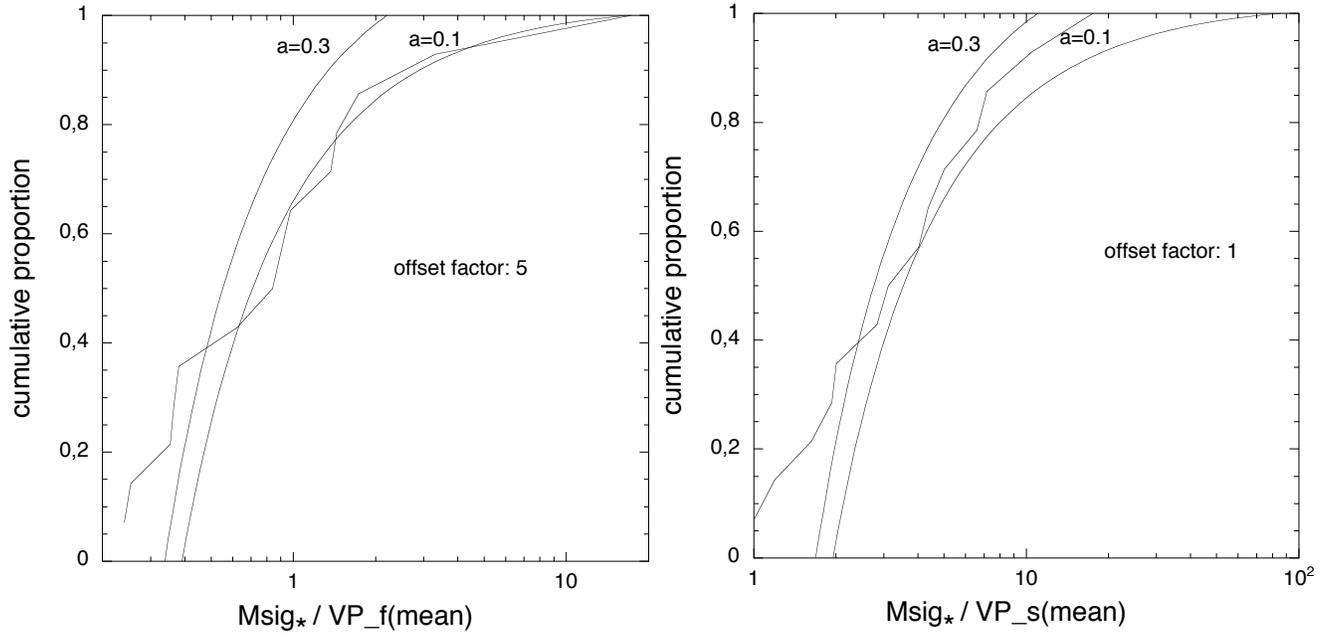}
\caption{Cumulative fraction of values of 
$M_{{\sigma*}}/{\rm VP}$ for the Onken sample of 14 
objects, compared to the theoretical distribution for 
two thick-disk models, as shown in Fig.\ 
\ref{fig-cum-VPdisk-on-VPisotrop}, for the FWHM-based values 
$M_{\sigma*}/{\rm VP_f}$ (left)
and the $\sigma_{\rm line}$-based values 
$M_{\sigma*}/{\rm VP_s}$ (right). The ``offset factor'' 
is described in the text.}
\label{fig-cum-MsigstaroVP}
\end{center}
\end{figure*}

In Fig.\ \ref{fig-cum-MsigstaroVP}, we show
the cumulative distribution of the values 
$M_{\sigma*}/{\rm VP}$ for the FWHM-based values of the 
virial product ${\rm VP_f}$ (left) and the $\sigma_{\rm line}$-based values
${\rm VP_s}$ (right) for the Onken sample of 14 
objects\footnote{Note that for the sake of simplicity, we will restrict 
the discussion only to quantities derived from the mean 
spectra since these do not differ significantly from 
those derived from the rms spectra. Also, in the cases 
of sources for which multiple datasets are available, 
we use an average value of VP so that each source 
is counted only once.}.
The ``offset factor'' in this and subsequent figures is 
the number by which the theoretical 
$M_{\sigma*}/{\rm VP}$ ratio has been divided to aid in 
comparison of the two distributions.  It is related to,
but is not identical to, the scale factor $f$ if the 
inclination is the sole cause of the discrepancies 
between $M_{\rm rev}$ and $M_{\sigma*}$,
but is otherwise completely arbitrary.
Fig.\ \ref{fig-cum-MsigstaroVP} shows that the $\sigma_{\rm 
line}$- based values $M_{\sigma*}/{\rm VP_s}$ do not match the theoretical distribution, while on the contrary
the FWHM-based values ${\rm VP_f}$ seem to  
match the theoretical distribution rather well at large 
values of $M_{\sigma*}/{\rm VP_f}$ for $a=0.1$. This suggests that 
the thick-disk BLR model probably has some merit, 
particularly
in describing the line core (to which FWHM is more 
sensitive), implying that FWHM has some dependence 
on inclination. However, the poor match of the
theoretical and observed distributions based on $\sigma_{\rm line}$ implies that $\sigma_{\rm line}$ is insensitive to source inclination. We speculate further in \S{\ref{section:discussion}}.

\begin{figure}
\begin{center}
\includegraphics[width=\linewidth]{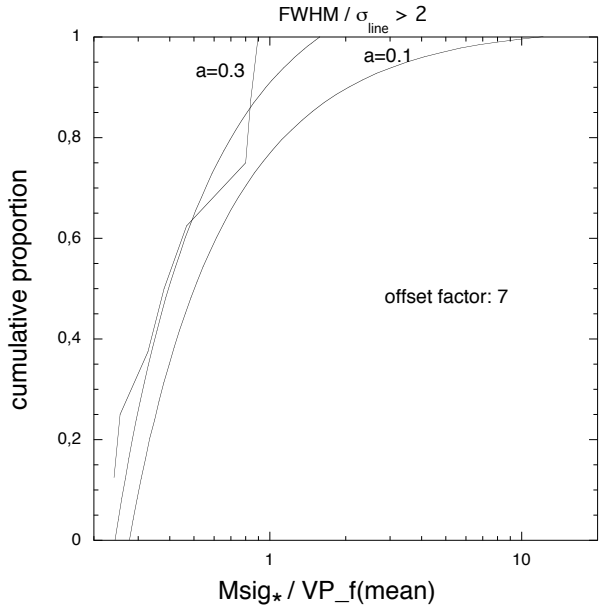}
\caption{Cumulative fraction of $M_{\sigma*}/{\rm VP_f}$  
for Population 2 objects (expanded to include all 
objects with
${\rm FWHM}/\sigma_{\rm line} > 2$).}
\label{fig-cum-MsigstaroVP_f-Pop2}
\end{center}
\end{figure}

We now consider separately Populations 1 and 2. 
This unfortunately exacerbates the problems associated 
with small-number statistics, but we find some 
important differences between the two populations that 
we believe are enlightening.
Figure \ref{fig-cum-MsigstaroVP_f-Pop2} shows the 
cumulative distribution of $M_{\sigma*}/{\rm VP_f}$ for
Population 2, although we have actually relaxed our 
orginal arbitrary criterion 
${\rm FWHM}/\sigma_{\rm line} > 2.35$ to 
${\rm FWHM}/\sigma_{\rm line} > 2$ in order to increase 
the sample size. Figure \ref{fig-cum-MsigstaroVP_f-Pop2} 
shows that the cumulative distribution is not 
well-described by the theoretical curve 
at large values of $M_{\sigma*}/{\rm VP_f}$. We conclude 
that our
generalized thick-disk model with $i < 45^\circ$ is not 
a good
description for this population. More specifically, the 
distribution of $M_{\sigma*}/{\rm VP_f}$ does not appear 
to be controlled by
inclination angle for this population.

\begin{figure*}
\begin{center}
\includegraphics[width=\linewidth]{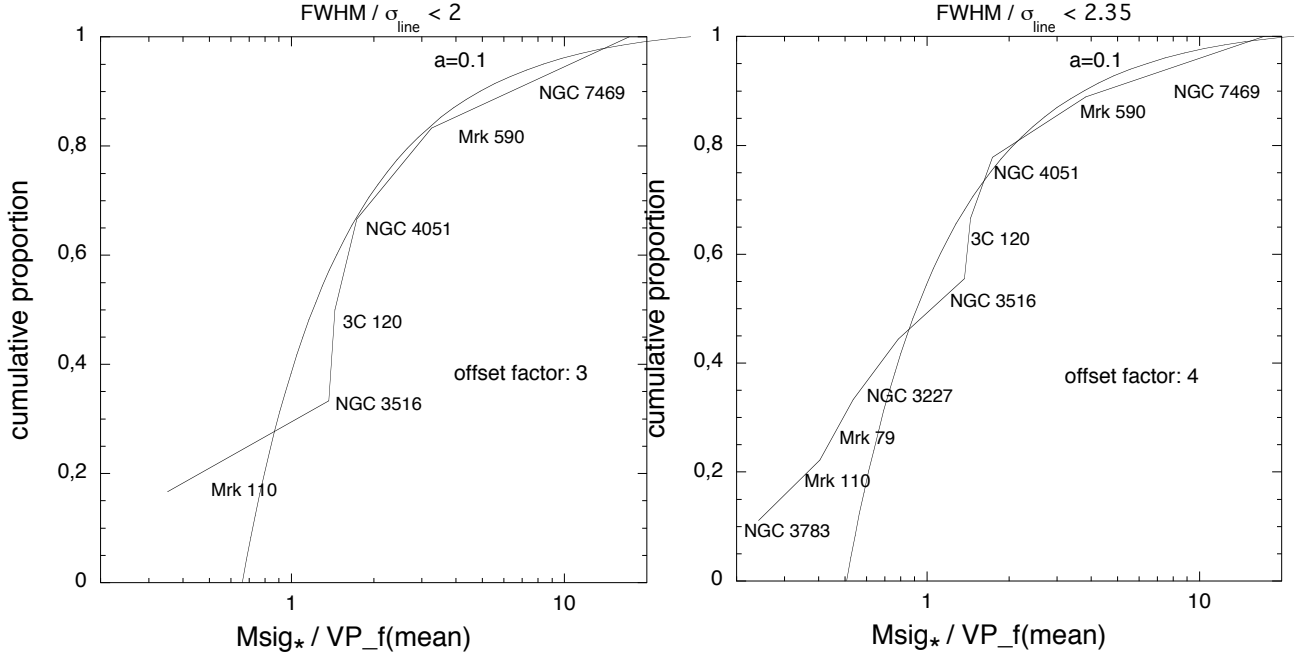}
\caption{Cumulative fraction $M_{\sigma*}/{\rm VP_f}$  
for Population 1 AGNs in the Onken sample, compared to 
the predicted distribution
for our generalized thick-disk model with $a = 0.1$.
In the left panel, Population 1 is restricted to AGNs 
with  ${\rm FWHM}/\sigma_{\rm line} < 2$, while in the right 
panel,  AGNs with
$2 < {\rm FWHM}/\sigma_{\rm line} < 2.35$ are also 
included.}
\label{fig-cum-MsigstaroVP_f-Pop1}
\end{center}
\end{figure*}

In contrast, the cumulative distribution of Population 1 
shown in Fig.\ \ref{fig-cum-MsigstaroVP_f-Pop1} is 
well-described by the theoretical curve at large values 
of $M_{\sigma*}/{\rm VP_f}$ for the case 
$a=0.1$. This figure shows clearly that the few AGNs 
with 
${\rm FWHM}/\sigma_{\rm line} < 2 $ seem to fit the 
theoretical distribution very well, but the AGNs with
$2 < {\rm FWHM}/\sigma_{\rm line} < 2.35 $ match the 
theoretical distribution rather less well.

We conclude that 
{\it among the sample of AGNs with small  
${\rm FWHM}/\sigma_{\rm line}$ ratios, the three objects
with the lowest values of $M_{\sigma*}/{\rm VP_f}$ 
{\rm (NGC 4051, Mrk 590 and NGC 7469)} are 
probably actually observed at low inclination.}
Although the statistics are very poor, we are led to the 
conclusion that {\it the difference in scale factor 
$f$ (Table 2)
between Population 1/A and Population 2/B is due at 
least partly to an inclination effect}. According to 
Fig.\ \ref{fig-cum-VPdisk-on-VPisotrop}, and taking 
into account the offset factor, {\it these three 
objects are probably
at inclinations $i \la 20{^\circ}$, and their masses 
$M_{\rm rev}$ could be underestimated by factors as 
large as an order of magnitude}.

While some properties of NLS1s can be attributed to either high
Eddington ratios or low inclination, a clear consensus based on the
preponderance of evidence is that accretion rate is probably the key
factor that defines the NLS1 class (see Boller 2000). While we agree
that some properties of NLS1s, in particular their rapid and large
amplitude X-ray variability, steep X-ray spectra, and the purported
larger fraction of bars in their host galaxies (Crenshaw, Kraemer, \&
Gabel 2003; Ohta, Aoki, \& Kawaguchi, in preparation)  can certainly not be
explained {\it only} by the inclination, we argue that inclination
does play an important, although not the dominant, role in defining this
class of object. In this context, we point out that 
Williams, Mathur, \& Pogge (2004) show that not all 
{\em optically selected} NLS1s seem to be high accretion sources;
many have characteristics more typical of broad-line objects.
These may well be the sources for which inclination is important.

\subsection{Inclinations of Individual Objects}

Unfortunately, there is currently no way to measure the inclination of
any AGN BLR to any reasonable accuracy, and there are few observables
that are even good indicators of inclination. Of the seven Population
2 objects in the Onken sample, estimates are available for only two
AGNs, 3C~390.3 and NGC~4151.  The former is one of the relatively rare
strongly double-peaked emitters; the double peaks are widely regarded
as the signature of an inclined disk in Keplerian rotation. Eracleous
et al.\ (1996), based on several indicators, conclude that the
inclination of the BLR in this system must be in the range $19 < i <
42{^\circ}$, and based on the profile fitting alone, Eracleous \&
Halpern (1994) find that $i = 24-30^{\circ}$.
At the very least, this relatively large inclination is consistent
with our finding that 3C~390.3 is not a member of Population 1.  On
the other hand, it is quite natural to find a relatively large
inclination for a powerful FRII radio source like 3C~390.3 that
does not show superluminal motion.  In the case of NGC 4151, the
deduced inclination ranges from $12$--$21{^\circ}$ (Boksenberg et al.\
1995) and 18--$23{^\circ}$ (Winge et al.\ 1999), to as high as
$40{^\circ}$ (Pedlar et al.\ 1998) or even $70^{\circ}$ (Kaiser et
al.\ 2000). There is thus no clear consensus on the inclination of
this system.

There are also few inclination estimates available for the Population
1 objects. The radio source 3C 120 has a superluminal jet, and
therefore must be seen fairly close to face-on. Marscher et al.\
(2002) estimate that the inclination of this source must be less than
$20{^\circ}$, which is consistent with the fact that it is a
Population 1 object and with Fig.\ \ref{fig-cum-MsigstaroVP_f-Pop1}.
In the case of Mrk 110, Kollatschny (2003b) showed that the variable
part of the broad lines of this bright NLS1 nucleus are redshifted with
respect to the systemic velocity, from which he deduced a
``gravitational mass'' of $1.4\times10^8\,M_{\odot}$, larger than the
``isotropic'' reverberation mass by almost one order of magnitude, and
requiring an inclination angle $ i \approx 19{^\circ}$. 
The position of Mrk 110 in 
Fig.\ \ref{fig-cum-MsigstaroVP_f-Pop1} indicates a relatively
large inclination angle, but we have seen that the objects located in
this part of the diagram do not appear to be very sensitive to
inclination, and are therefore subject to large uncertainties in the
determination of the inclination.

Popovi\'c et al.\ (2004) attempted to determine inclinations for
individual objects by detailed fitting of individual line profiles.
Unfortunately, their model has more free parameters than observational
constraints and these authors are obliged to impose at least one arbitrary
constraint, which they choose to be the emissivity as a function of
the radius.  They generally find quite small inclinations, but with
large uncertainties, and inescapable model dependence.

\subsection{The Influence of Inclination on the Scale 
Factor}

The similarity between the offset factor of 3 found for the
Population1/A objects in Fig.\ \ref{fig-cum-MsigstaroVP_f-Pop1} and
the mass scale factor of 2 to 2.5 of this population (Table \ref{table2})
could give the impression that the scale factor is determined mainly
by inclination, but this is not the case.  If indeed the inclination
effect were the principal factor affecting the scale factor $f$, then
the average ratio would be 
\begin{equation}
\langle \frac{{\rm VP_{thick\ disk}}}{{\rm VP_{isotropic}}}
\rangle ={ \int{} A(i)\ \sin i \ di \over 1 - \cos i_0}.
\label{eq-scale1}
\end{equation}
Using eq.\ (\ref{eq-G}), we obtain
\begin{eqnarray}
\langle {{\rm VP}_{\rm thick\ disk}\over {\rm VP}_{\rm isotropic}}
\rangle & = & {1\over 2b \left(1 - \cos i_0\right)} \\
\nonumber
&&\times \left[ \ln \left({b+1\over b+\cos i_0 }\right) 
+
\ln \left( { b - \cos i_0 \over b-1}\right) \right],
\label{eq-scale2}
\end{eqnarray}
where $b=(1+\langle a \rangle^2)^{1/2}$ and
$\langle a \rangle$ is the inclination averaged value of $a$.
For $\langle a \rangle =0.1$ and $i_0=45^{\circ}$, we 
find
$\langle {\rm VP}_{\rm thick\ disk}/{\rm VP}_{\rm 
isotropic}\rangle = 7.1$ (or 4.8 for $a=0.2$). This is 
larger than the scale factor for  Population 1/A, which 
means that while the inclination does certainly play a 
role in the determining the scale factor $f$, it is not 
the  dominant source of the differences between 
$M_{\rm rev}$ and $M_{\sigma*}$.

\section{Discussion}
\label{section:discussion}

An obvious question to ask is why the FWHM would be 
more dependent on the inclination than the line 
dispersion $\sigma_{\rm line}$?
While a definitive answer is not possible, we speculate 
that the line wings, to which $\sigma_{\rm line}$ is 
relatively more sensitive, arise primarily in a 
more-or-less isotropic component, perhaps in the form of a 
``disk wind.'' The line core, to which FWHM is more 
sensitive, might then arise primarily in a Keplerian 
disk component, and thus FWHM would be more 
sensitive to inclination.
Such a scenario would also account, at least in part,
for the smaller values of ${\rm FWHM}/\sigma_{\rm 
line}$ ratios in Population 1 objects, as the presence 
of stronger winds in Population 1
would naturally correspond to higher Eddington ratios 
than Population 2. By analogy, we know indeed that hot 
stars radiating close to their Eddington limit have 
strong winds, and there seems to be an emerging 
consensus that this is also the case for quasars 
accreting at a high Eddington ratio
(King \& Pounds 2003, Pounds et al.\ 2003).

We can further speculate as to how the BLR can differ 
between low and high Eddington ratio objects. At large 
distances from the center, the disk is self-gravitating 
and gravitationally unstable (see Collin \& Hur\'e 
2001).  As a consequence of the gravitational 
instability, the disk should be broken into clumps. The 
fate of these clumps, and more generally the state of 
the disk in this region, the accretion mechanism, and 
the way angular momentum is removed, are unknown 
(e.g., Collin \& Kawaguchi 2004), but we might 
speculate that the ``disk'' would be made of discrete 
clumps and it seems natural to identify these clumps 
with the BLR clouds.  The heating of the cloud system 
would be thus provided by the collisions between the 
clouds, as suggested also for the molecular torus 
(Krolik \& Begelman 1988). The larger the gravitational 
instability, the larger the heating rate (cf. Lodato \& 
Rice 2004).  It is also probable that a fraction of the 
clumps constitute the basis of a wind, which would be 
more efficient when the gravitational instability is 
strong.

\begin{figure}
\begin{center}
\includegraphics[width=\linewidth]{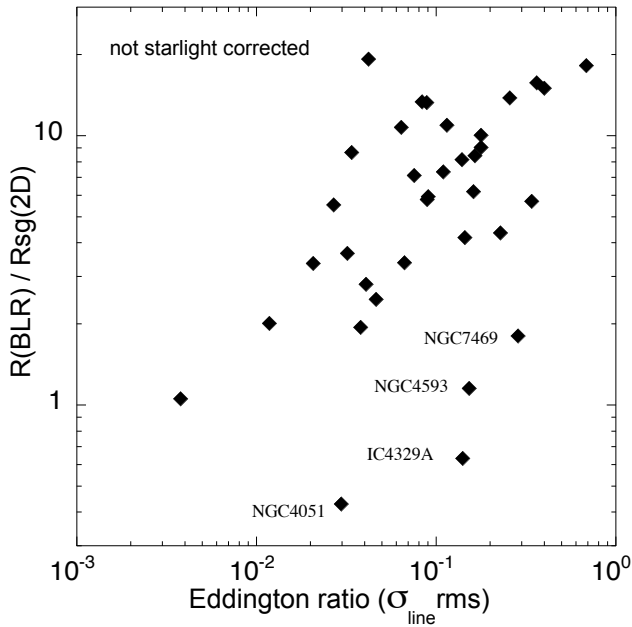}
\caption{The ratio $R_{\rm BLR}/R_{\rm sg}$, as 
described in the text. The four outliers are objects 
whose luminosity and Eddington ratio are probably 
overestimated (see the text). If $R_{\rm BLR}/R_{\rm sg} > 4$, the BLR 
is gravitationally unstable.}
\label{fig-RBLRsRsg-vs-Redd}
\end{center}
\end{figure}

Figure \ref{fig-RBLRsRsg-vs-Redd} displays the ratios 
$R_{\rm BLR}/R_{\rm sg}$ versus the Eddington ratio for 
the reverberation-mapped sample, where $R_{\rm sg}$ is 
the radius above which the 
self-gravity of the disk overcomes the vertical 
component of the central gravity. Here $R_{\rm sg}$  
has been computed with a 2D simulation using real 
opacities (Hur\'e 1998)\footnote{In the course of these 
computations, we have confirmed that the analytical 
formulae for $R_{\rm sg}$ given by Kawaguchi et al.\ 
(2004a) show consistent results (with systematically 
larger values by a factor of $\sim 1.5$).} and assuming 
a viscosity coefficient $\alpha=0.1$. There is a clear 
correlation between $R_{\rm BLR}/R_{\rm sg}$ and the 
Eddington ratio, with the exception of the four 
outliers labeled in the Figure; three are among the least-luminous AGNs in the sample, so their luminosity could have been overestimated as mentioned in \S 2, and NGC 7469 is the object suspected to have the largest inclination effect, so its mass could be strongly underestimated, and thus its Eddington ratio be also overestimated. The 
radius at which the disk becomes gravitationnally 
unstable is about four times larger than $R_{\rm sg}$. 
Thus, according to Fig.\ \ref{fig-RBLRsRsg-vs-Redd},  
the BLR lies in the gravitationnally unstable region in 
the objects with $L/L_{\rm Edd} \ga 0.1$. On the other 
hand, objects with 
low Eddington ratios ($L/L_{\rm Edd} \la 0.03$) have 
their BLRs in the gravitationnally stable region. We 
therefore speculate that the BLR of Population 1/A 
AGNs, which is gravitationally unstable, is more 
influenced by the wind than Population 2/B, thus 
explaining the difference in the line profiles and the 
difference in the 
${\rm FWHM}/\sigma_{\rm line}$ ratio.

Another possible explanation for
why the FWHM/$\sigma_{\rm line}$
ratio is larger in Population 2/B objects
is that the structure of the inner accretion disk is
different in these objects. Population 2/B
includes objects like the broad-line radio galaxy
3C~390.3, which have strongly
double-peaked line profiles. Chen \& Halpern (1989) have suggested
that the inner disk is an inflated ion-supported torus which
illuminates the outer line-emitting part of the extended disk. 
It is indeed 
thought that a hot advection dominated accretion flow (ADAF,
see Narayan \& Yi 1994 and subsequent works), or more
generally a radiatively inefficient accretion flow (RIAF), is
present close to the BH, and that such structures become
increasingly prominent with decreasing Eddington ratio,
If the Eddington ratio is
small, we  expect that an annular region at relatively small
distance from the BH is heated by the X-rays from the
geometrically thick RIAF, giving rise to very broad two-peaked
line profiles. Note however that in this case, 
the line-emitting region would be a thin disk,
which is very sensitive to the
inclination, contrary to what we have deduced previously.

Finally, Murray \& Chiang (1997) argue that a varying
   optical depth in outflowing disk winds can explain the presence of
   single or double peaked line profiles. In this scenario, a low
   optical depth would tend to generate double-peaked lines while
   single-peaked profiles are the result of high optical depths
   in the wind.

Could a combination of these phenomena (gravitational
instability, inflated inner hot disk, wind) which are all
linked with the Eddington ratio, combined with the influence
of inclination, explain the variation of the 
${\rm FWHM}/\sigma_{\rm line}$ ratio among the AGN population? 
Obviously it will be
necessary to consider larger samples of objects for which
these two parameters are available in order to check these
different ideas.

\section{Summary}

In this contribution, we have initiated a study
of the relationship between AGN BH masses
and other physical properties of AGNs that can be
discerned from  broad emission-line profiles. We have used
the ratio of FWHM to the line dispersion
$\sigma_{\rm line}$ to characterize the emission-line profiles
and have shown that this ratio is anticorrelated
with Eddington ratio and with line widths;
broader emission lines tend to have relatively flat-topped
profiles, and narrower lines have more extended wings.
We separate AGNs into two populations on
the basis of their H$\beta$ profiles,
a Population 1 with ${\rm FWHM}/\sigma_{\rm line} < 2.35$
and a Population 2 with ${\rm FWHM}/\sigma_{\rm line} > 2.35$.
Not surprisingly, these two populations overlap strongly
with Populations A and B of Sulentic et al.\ (2000) which
are separated by FWHM only.

We then make the assumption that AGNs follow the same
$M_{\rm BH}$--$\sigma_*$ relationship as quiescent galaxies
and scale the virial product, the observable parameter,
to determine the statistical value of the scaling factor
$\langle f \rangle$ of eq.\ (1) (cf.\ Onken et al.\ 2004). 
We do this for virial products
based on both $\sigma_{\rm line}$ and FWHM, as measured in
both the mean and rms spectra. We find that, to within
the uncertainties, the scaling factor is constant for
both Populations 1 and 2 for virial products based
on using $\sigma_{\rm line}$ as the line-width measure.
On the other hand, the scaling factors are significantly
different for the two populations if the virial
product is based on FWHM. This means that
$\sigma_{\rm line}$ is a {\em less biased} mass estimator
than is FWHM. However, we show that it is possible to remove
heuristically the bias in masses based on FWHM and obtain
masses estimates consistent with those based on 
$\sigma_{\rm line}$.

For the 14 objects with measured bulge velocity dispersions
$\sigma_*$, we have compared the black hole mass predicted by
the $M_{\rm BH}$--$\sigma_*$ relationship with the mass
determined by reverberation mapping. 
By comparing the distribution of the ratio of these masses with
the distribution expected from a generalized thick disk model
of the BLR, we find {\em statistical} evidence that in some small
fraction of cases, the reverberation-based
BH masses in Population 1 objects are underestimated on account
of inclination effects. We find no evidence for inclination
effects in Population 2 objects. We speculate that the difference
between the two populations is the relative strength of
a disk-wind component, which is stronger in Population 1.
Finally, we discuss the possible role of self-gravity as the physical
driver controlling the strength of the disk wind. We find
a stronger wind is expected for larger Eddington ratios,
which is consistent with the smaller
${\rm FWHM}/\sigma_{\rm line}$ ratios found for Population 1
objects.

\begin{acknowledgements}

We would like to thank Jean-Marc Hur\'{e} and
Didier Pelat for useful discussions.
T.K.\ is grateful for financial support through 
a JSPS Postdoctoral Fellowship.
This research was supported in part by the US National Science
Foundation through grant AST-0205964 (B.M.P.) and AST-0307384
(M.V.) and by NASA through
grant HST-AR-10691 from the Space Telescope Science Institute (M.V. and B.M.P.).

\end{acknowledgements}

\bigskip

\end{document}